\documentclass[aps,pre,reprint,superscriptaddress,amsmath,amssymb]{revtex4-1}

\usepackage{graphicx}
\usepackage{dcolumn}
\usepackage{bm}
\usepackage{hyperref}
\usepackage{comment}
\usepackage{color}
\usepackage{soul}

\usepackage{graphicx}
\usepackage{dcolumn}
\usepackage{bm}
\usepackage{color}
\usepackage{marvosym}
\usepackage{longtable}
\usepackage{booktabs}

\begin{document}

\preprint{APS/123-QED}

\title{Salt polygons and porous media convection}

\author{Jana Lasser}
\email{jana.lasser@tugraz.at}
\affiliation{Graz University of Technology, Institute for Interactive Systems and Data Science, Inffeldgasse 16c, 8010 Graz, Austria}
\affiliation{Max Planck Institute for Dynamics and Self-Organization Am Fassberg 17, 37077 G\"ottingen, Germany}

\author{Joanna M. Nield}
\affiliation{Geography and Environmental Science, University of Southampton Highfield, Southampton SO17 1BJ, UK}

\author{Marcel Ernst}
\affiliation{Max Planck Institute for Dynamics and Self-Organization Am Fassberg 17, 37077 G\"ottingen, Germany}

\author{Volker Karius}
\affiliation{Geowissenschaftliches Zentrum, Georg-August-University Goldschmidtstrasse 3, 37077 G\"ottingen, Germany}

\author{Giles F.S. Wiggs}
\affiliation{School of Geography and the Environment, University of Oxford, South Parks Road, Oxford OX1 3QY, UK}

\author{Matthew R. Threadgold}
\affiliation{School of Mathematics, University of Leeds, Leeds, LS2 9JT, UK}

\author{C\'edric Beaume}
\affiliation{School of Mathematics, University of Leeds, Leeds, LS2 9JT, UK}

\author{Lucas Goehring}
\email{lucas.goehring@ntu.ac.uk}
\affiliation{School of Science and Technology, Nottingham Trent University Nottingham NG11 8NS, UK}

\date{\today}

\begin{abstract}
From fairy circles to patterned ground and columnar joints, natural patterns spontaneously appear in many complex geophysical settings. Here, we investigate the origins of polygonally patterned crusts of salt playa and salt pans. These beautifully regular features, approximately a meter in diameter, are found worldwide and are fundamentally important to the transport of salt and dust in arid regions. We show that they are consistent with the surface expression of buoyancy-driven convection in the porous soil beneath a salt crust. By combining quantitative results from direct field observations, analogue experiments and numerical simulations, we further determine the conditions under which salt polygons should form, as well as how their characteristic size emerges.
\end{abstract}


\maketitle

\section{Introduction}
\begin{figure}
\centering
\includegraphics[width=0.47\textwidth]{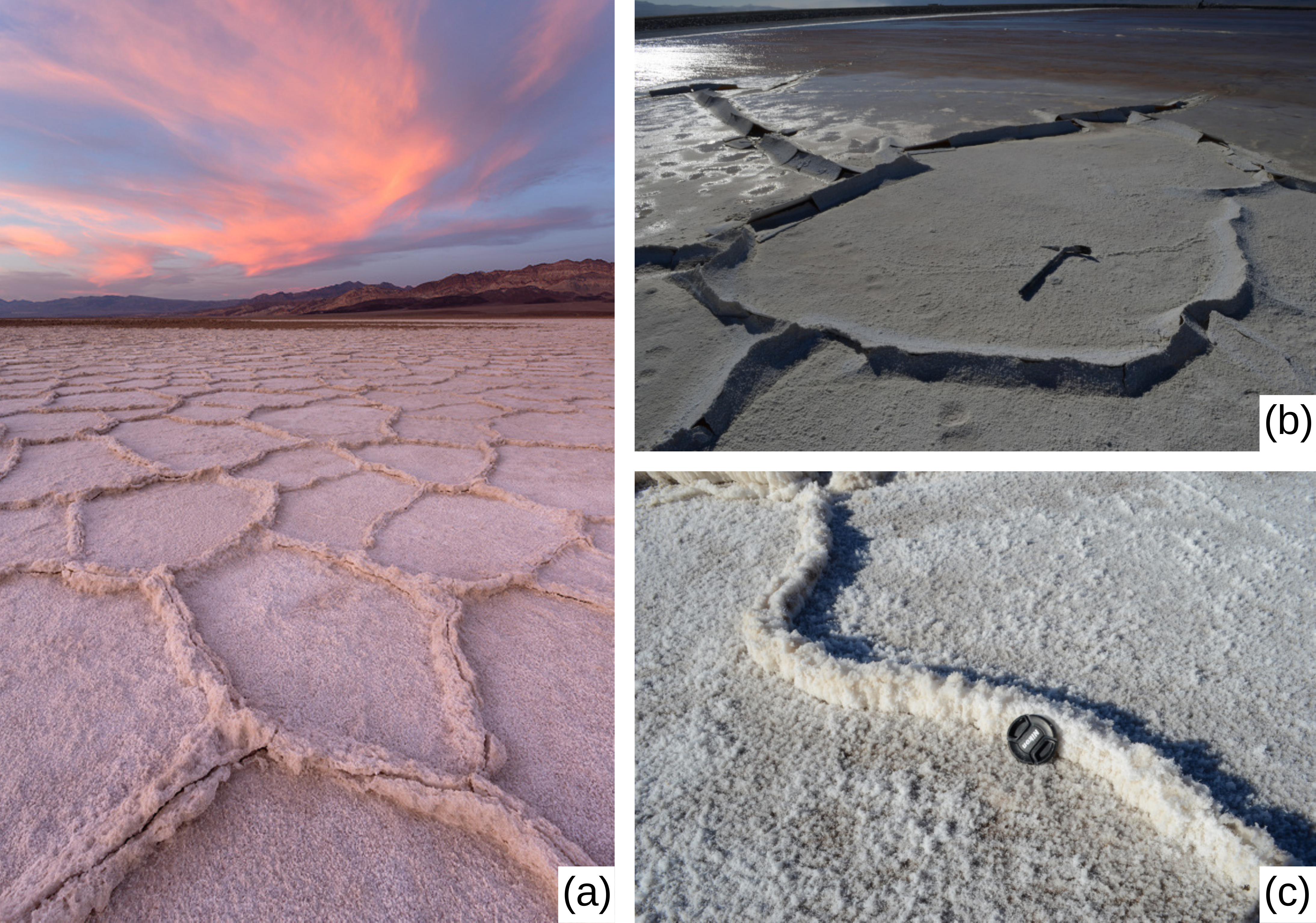}
\caption{\label{fig:fig1} Typical salt polygon patterns at \textbf{(a)}, \textbf{(c)} Badwater Basin in Death Valley and \textbf{(b)}, Owens Lake, California (image \textbf{(a)} courtesy Sarah Marino).}
\end{figure}

Salt deserts or dry salt lakes are amongst the most inhospitable places of our planet.  Their otherworldly shapes inspire the imagination (\textit{e.g.} Star Wars' desert planet Crait, or the million tourists annually visiting Death Valley \cite{DEVA}), and are an important drive of climate processes \cite{LOWENSTEIN1985,GILL1996,BREIRE2000,BRYANT2002,PROSPERO2002,Klose2019,LI2021}. Variously referred to as salt pans, playas, or dry lakes, the immediately prominent feature of such landscapes (Fig.~\ref{fig:fig1}) is a characteristic tiling of polygons, formed by ridges in the salt-encrusted surface. These patterns are seen around the world~\cite{ERICKSEN1978,WADGE1994,lasser2020surface,KRINSLEY1970,dang2018polygonal, SANFORD2001, NIELD2015, martinez2020morfologia}, including Salar de Uyuni in Chile~\cite{ERICKSEN1978}, Chott el Djerid in Tunisia \cite{WADGE1994}, Badwater Basin in California~\cite{lasser2020surface}, Dasht-e Kavir in Iran~\cite{KRINSLEY1970} and Dalangtan Playa in China~\cite{dang2018polygonal}.  Although they have been argued to share similarities with fracture \cite{KRINSLEY1970} or buckling \cite{CHRISTIANSEN1963} patterns, a quantitative and predictive mechanism for the emergence of the patterns remains obscure. 

A challenge to any explanation is that the polygonal patterns in salt crusts are remarkably similar wherever they occur, despite local differences in geology, salt chemistry, and environmental conditions.  For example, while crusts can vary from sub-centimeter to meters in thickness \cite{KRINSLEY1970, LOWENSTEIN1985, LOKIER2012} they predominantly express polygons about 1--2$\,$m across~\cite{lasser2020surface, KRINSLEY1970, martinez2020morfologia, CHRISTIANSEN1963, NIELD2015, LOWENSTEIN1985, DIXON2009, DEDECKKER1988, LOKIER2012}. The same patterns also appear for pure halite crusts such as those of Badwater Basin~\cite{lasser2020surface} and pans in Iran~\cite{KRINSLEY1970} as well as the 
sulfate-rich crusts found for example at Owens Lake (California)~\cite{lasser2020surface}, Sua Pan~\cite{ECKARDT2008}, the coastal sabkhas of Abu Dhabi~\cite{SANFORD2001} and Salar de Uyuni~\cite{ERICKSEN1978}.

The salt crusts of dry lakes are known to be dynamic over months to years~\cite{CHRISTIANSEN1963,LOWENSTEIN1985, LOKIER2012, NIELD2013, NIELD2015,Milewski2020}, and couple to other environmental processes.  {They form in areas with a negative water balance (\textit{i.e.} where evaporation significantly exceeds precipitation) and the dominant water balance is between a perennial groundwater flow seeping in from distant sources, and the evaporation of this water from the surface of the dry lake \cite{LOWENSTEIN1985,BREIRE2000,Reynolds2007}.  The groundwater brings dissolved salts with it, which accumulate at the surface in a crust of evaporites.} Wind blowing over the crust entrains dust, which forms a significant proportion of the global atmospheric dust production \cite{GILL1996,PROSPERO2002,Klose2019} and of mineral transport to the oceans~\cite{FUNG2000}.  As one example, dust from the dry Owens Lake has posed health problems for people living nearby \cite{CAHILL1996,GILL2002}, and the site is the center of a decades-long, intense remediation effort~\cite{OwensLakeDust}; {a preferred method is the use of shallow flooding to encourage formation of a continuous, active surface crust~\cite{GROENEVELD2010,GROENEVELD2013}.}

{While playas are inherently associated with dust sources, these sources are temporally and spatially variable~\cite{Reynolds2007,Tollerud2014,BADDOCK2016,Sweeney2016}, and there is a lack of understanding of the relevant processes of how crusts form, are sustained, and degrade~\cite{BRYANT2013,Klose2019}.  The height of polygon ridges affects the threshold wind speed for dust entrainment~\cite{NIELD2016,Klose2019} and complete coverage of a crust reduces the sediment available for dust emission compared to a patchy crust.  In order to better account for these complexities we need to understand the relationships between surface and sub-surface drivers of playa dust sources~\cite{Tollerud2014,Klose2019}.}


Salt crusts also modify evaporation and heat flux from the playa surface~\cite{BRYANT2002}, and hence the water and energy balances of these fragile environments. {In particular, experimental and field research consistently show that continuous crusts reduce evaporation rates on playa surfaces by an order of magnitude compared to climate-derived estimates~\cite{ELOUKABI2013,NIELD2016,LI2021,TYLER1997,GROENEVELD2010}.  As a result, the observed groundwater evaporation rates at sites with active and continuous salt crusts tend to lie in the relatively low range of 0.2--0.7 mm/day, including seasonal variations~\cite{TYLER1997,GROENEVELD2010,DEMEO2003,BRUNNER2004,KAMPF2005,WOOD2002}.} 

\begin{figure}
\centering
\includegraphics[width=0.47\textwidth]{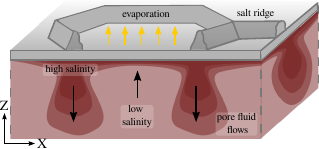}
\caption{\label{fig:fig2} Proposed dynamics of patterned salt crusts. The dominant fluid motions are shown by the black arrows, and the water salinity is indicated by the color contours.}
\end{figure}

Research on salt pans has typically focused on either the dynamics of their complex subsurface flows~\cite{WOODING1960a,HOMSY1976,WOODING1997b, WOODING2007,VANDAM2009} or their crusts \cite{CHRISTIANSEN1963,LOKIER2012,NIELD2015,NIELD2016}, without considering how these features interact.  Here we show that by treating the surface of a salt playa together with the fluid in the porous media of the soil near its surface, the origins, dynamics, length-scale and shape of the polygonal patterns in salt crusts can be apprehended.  {As we develop and test such a model of salt crust formation, we will show how it makes explicit predictions of crust growth rates, the conditions necessary to form ridges and continuous crusts, and other inputs into dust or groundwater models of these extreme environments.} In particular, by combining theoretical analysis, numerical simulations, experiments and field observations, we will demonstrate how the density-driven convective dynamics of groundwater can lead to variations in the salt flux into the crust, with faster precipitation along the boundaries between convective cells facilitating the growth of a polygonal network of salt ridges (see Fig.~\ref{fig:fig2}).  Under measured field conditions the size of the expected convective cells matches the scale observed for the surface crust patterns and makes accurate predictions about the crust growth rates. Finally, we will present direct field-based evidence of convective dynamics that are correlated to the surface crust patterns.

\section{Phenomenology of salt polygons}\label{sec:phenomenology_of_salt_polygons}

Any model for the emergence of salt crust patterns should be able to convincingly explain their commonly observed features. {Polygonal networks are one of the more common emergent patterns in geophysics, seen for columnar jointing~\cite{GOEHRING2009}, ice-wedge polygons~\cite{LACHENBRUCH1962,SLETTEN2003}, polygonal terrain~\cite{Kessler2003} some types of mud-cracks~\cite{GOEHRING2013} as well as convection~\cite{FU2013,Paoli2022} for example.  As such, a suite of more detailed or quantitative predictions are important to evaluate any model that displays polygonal features.}  

Based on the known phenomenology of salt playa, the most significant tests for a predictive model include: (a) The driving mechanism should be specific to the geophysical conditions encountered in salt lakes~\cite{KRINSLEY1970, SANFORD2001, ERICKSEN1978, lasser2020surface, NIELD2015, martinez2020morfologia, WADGE1994} and (b) should spatially modulate salt transport, leading to increased salt precipitation at ridges in the crust~\cite{KRINSLEY1970, FRYBERGER1983, LOWENSTEIN1985}; (c) The response should be fast enough, and stable enough, to account for ridge growth on a scale of months~\cite{LOWENSTEIN1985, NIELD2013, NIELD2015, LOKIER2012}; (d) The model dynamics should generate closed polygonal patterns of {relatively narrow} ridges spaced a few meters apart~\cite{CHRISTIANSEN1963, KRINSLEY1970, NIELD2015, LOKIER2012, lasser2020surface, LOKIER2012, martinez2020morfologia};  and (e) the characteristics and scale of the patterns should be robust to environmental variations, including different types of soil ranging from silt to sand~\cite{lasser2020surface}, salt compositions ranging from pure halite crusts to those with significant amounts of sulphates and carbonates (\textit{e.g.} trona)~\cite{lasser2020surface, ERICKSEN1978, SANFORD2001, KRINSLEY1970, TYLER1997} and any local differences in climate and variables like the evaporation rate~\cite{GROENEVELD2010, TYLER1997, DEMEO2003, BRUNNER2004}. 

Previous models of salt polygons have mainly considered mechanical effects.  The earliest argument was made by Christiansen in 1963~\cite{CHRISTIANSEN1963}.  He suggested that the shapes were the result of compressive stress, due to salt precipitation and temperature changes, which caused the crust to first buckle and then crack as the growing ridges thrust upwards.  He noted that the spacing of the buckling features should be proportional to the thickness of the stressed crust and also depend on its strength; these predictions are consistent with the modern understanding of wrinkling phenomena~\cite{BOWDEN1998,MAHA2005,LI2012}.  However,  wrinkles due to isotropic compression typically form parallel features, such as stripes, rather than closed shapes~\cite{BOWDEN1998,MAHA2005,LI2012}; polygons can form by compression, but only by folds into an elastic layer, rather than ridges projecting out of a layer~\cite{DERVAUX2012,LI2012}.  A more fracture-focused model was proposed by Krinsley~\cite{KRINSLEY1970} in his 1970 survey of Iranian salt pans.  He argued that desiccation places the crust into tension, rather than compression, which is relieved by fracture.  The cracks then allow for groundwater to more easily reach the surface to evaporate, so that ridges grow up along them.  As the crust develops faster at the ridges, these can buckle or thrust upwards (as in Fig.~\ref{fig:fig1}(b)).  Like wrinkling, for cracks the expected feature size is a few times the thickness of the cracking layer~\cite{LACHENBRUCH1961, SHORLIN2000}.  

Both the wrinkling~\cite{FRYBERGER1983,LOWENSTEIN1985} and fracture~\cite{DIXON2009, TUCKER1981, DEDECKKER1988, LOKIER2012} models can explain the qualitative appearance of cracks and thrusting at the ridges in salt crusts.  The main difficulty with any purely mechanical model, however, is length-scale selection.  The natural spacing between buckles or cracks is proportional to the thickness of the layer under stress~\cite{LACHENBRUCH1961,SHORLIN2000,BOWDEN1998,MAHA2005,DERVAUX2012,LI2012}.  As some geophysical examples, the scale of ice-wedge polygons in permafrost is determined by the depth at which annual temperature changes diffuse into the ground \cite{LACHENBRUCH1962,MELLON2013}, whereas the size of columnar joints in lava records how it cooled, and in particular the thickness of a cooling front invading into an initially molten lava formation \cite{GOEHRING2009}.  For salt playa, crust thicknesses vary widely, so cannot explain the robust polygonal scale seen in nature, and no other convincing explanation for the feature spacing has been proposed. 

As an important mechanism one may also consider the crust composition itself, which can include minerals in different hydration states.  For example, thenardite (Na$_2$SO$_4$) transforms into mirabilite (Na$_2$SO$_4\cdot10$H$_2$O), and expands its volume by $\sim$320\%, at temperatures easily seen in salt deserts~\cite{steiger2008crystallization}.  While phase changes can undoubtably generate stresses, the difficulty of length-scale selection remains, given similar patterns in crusts of different thickness.  In addition, salt polygons are routinely seen in crusts with almost pure halite (NaCl) composition~\cite{lasser2020surface,KRINSLEY1970}, where such effects are absent.

\section{Buoyancy-driven convection as a mechanism for pattern formation}\label{sec:buoyancy_driven_convection_as_a_mechanism_for_pattern_formation}

As discussed, explanations for the formation of salt polygons have so far focused on mechanical effects.  Crust growth is implied in these cases, but the salt source feeding this growth is not modeled explicitly.  In other contexts there is a well-developed literature regarding buoyancy-driven convection in porous media, including salt lakes~\cite{WOODING1997b,WOODING2007,LASSER2021}, sabkhas~\cite{STEVENS2009,VANDAM2009}, carbon sequestration applications~\cite{NEUFELD2010,FU2013,SLIM2014} and thermally-driven cases~\cite{WOODING1960a,HEWITT2014,Paoli2022}; the model we develop here builds on a recent study of the linear stability and approach to a statistically steady state of convection in a dry salt lake \cite{LASSER2021}.  However, the effects of such flows on a salt crust have not yet been explored.  Here, we will examine the implications of convection in the soil beneath a salt crust and its predictions for the emergence of salt polygons.  

We consider a coupling of surface salt patterns to subsurface flows, as visualized in Fig.~\ref{fig:fig2}.  Briefly, surface evaporation will leave the near-surface groundwater enriched in salt, and heavier than the fluid beneath it. This can lead to precipitation of salt at the surface as well as convection in the soil, with narrow, regularly spaced downwellings of high salinity \cite{LASSER2021}.  We will show that convection should result in a higher salt flux into the surface above the downwellings, where salinity gradients in the groundwater are weaker. We argue that this preferential precipitation of salt templates the growth of ridges at the locations of the downwellings. At upwelling plumes the larger salinity gradients strengthen the diffusive transport away from the surface, rather than advective transport towards it, and crust growth is slower.

Our main field site, Owens Lake (Fig.~1(b)), has a typical crust pattern in a well-studied and controlled landscape, and can introduce our modeling assumptions. This dry, terminal saline lake has an aquifer extending from the near-surface to over 150$\,$m depth \cite{GULER2004}. The groundwater carries dissolved salts \cite{TYLER1997,RYU2006,GILL2002}, which collect in an evaporite pan of about 300 km$^2$ \cite{GULER2004,GILL1996} and which are particularly concentrated in the fluid within a meter or so of the surface \cite{TYLER1997,lasser2020surface}. {Average annual groundwater evaporation rates of $E = 0.4 \pm 0.1 $ mm/day have been measured in areas with active salt crusts~\cite{TYLER1997}.  The near-surface soil is a silt to fine sand with high porosity of $\varphi = 0.7$~\cite{TYLER1997}.  A tortuosity-corrected diffusion constant, \mbox{$D=1.00\pm 0.24 \times10^{-9}\,$m$^2$\,s$^{-1}$} is representative of the mix of dissolved salts present at the site (mainly NaCl and Na$_2$SO$_4$).} Efforts to control dust emission from the lake involve shallow flooding \cite{GROENEVELD2013}, vegetation \cite{LANCASTER1997}, gravel cover and encouraging crust growth \cite{GROENEVELD2010}. As shown in supplemental movie S1, after a shallow-flooding event the soil is saturated with water, which evaporates and leaves behind salts that crystallize into a continuous crust, covered by a network of slowly growing ridges.  {Given the high quality of the observational record at Owens Lake, and in order to make quantitative predictions relevant to the remediation of this important site, we will use representative values from it throughout this section.}

\subsection{Governing equations}
The transport of fluid beneath a dry salt lake can be treated as a Darcy flow in a porous medium, with evaporation occurring at the surface and being fed from below by water with some background salinity (see \textit{e.g.} \cite{WOODING1960a, WOODING1997b,VANDUJIN2002,LASSER2021}).  Such a system will naturally develop a salinity gradient below the surface, which can become unstable to convective overturning \cite{WOODING1960a,WOODING1997b,LASSER2021}. 

As a prototype case we consider a large, flat salt lake with an average surface evaporation rate $E$, and where the groundwater is recharged from some deep, distant reservoir. For the subsurface flows we model the volumetric flux, or superficial fluid velocity, $\mathbf{q}$, of a fluid of pressure $p$ moving through a porous medium of constant uniform porosity $\varphi$ and permeability $\kappa$.  By mass conservation $E$ is also the average of the upward component of the fluid flux, $q_z$, at any depth as well as the recharge rate from the reservoir.   The fluid has a viscosity $\mu$ and carries dissolved salt, whose diffusivity $D$ can be corrected for the presence of different ions as well as tortuosity \cite{BOURDEAU1996,BOUDREAU2011}.

The dissolved salt contributes to the density of the fluid, $\rho$, and hence to buoyancy forces. In salt lakes, thermal buoyancy is several orders of magnitude weaker than solutal buoyancy, and can be neglected \cite{WOODING1997b,LASSER2021}. 
Using the Boussinesq approximation, $\rho = \rho_0 +S\Delta \rho$, where $\rho_0$ is the density of the reservoir fluid (including its dissolved salts, see Section~\ref{subsec:salt_flux_and_crust_growth}), and $\rho_1 = \rho_0 + \Delta\rho$ is the density of a saturated solution.   The relative salinity $S$ mediates between these limits: $S=0$ corresponds to the background salinity of fluid entering from a distant source ($z\rightarrow -\infty$), whereas $S=1$ represents a salt-saturated brine in contact with the crust (at $z=0$).

The governing equations are then a continuity equation for incompressible fluid flow, an advection-diffusion equation for the relative salinity, and Darcy's law:
\begin{eqnarray}
	\bm{\nabla} \boldsymbol{\cdot} \mathbf{q} &=& 0, \label{dim1}\\
	\varphi\frac{\partial S}{\partial t} &=&  \varphi D\nabla^2 S - \mathbf{q} \boldsymbol{\cdot} \nabla S,\label{dim2}\\
	\mathbf{q} &=& -\frac{\kappa}{\mu}\left( \nabla p + \rho g\hat{\bm{z}}\right),\label{dim3} 
\end{eqnarray}
where $g$ is the acceleration due to gravity, and $\hat{\bm{z}}$ is an upward-pointing unit vector. These mass and momentum balances can describe a variety of porous media flows, including solutal flows in playas and sabkhas \cite{WOODING1997b,BOUFADEL1999,VANDUJIN2002,WOODING2007,LASSER2021}, CO$_2$-rich flows in carbon sequestration applications \cite{SLIM2014,LOODTS2014} and, with an appropriate transformation of variables, flows driven by thermal buoyancy~\cite{WOODING1960a,ELDER1967,HEWITT2014}. A review of these phenomena was recently given in Ref. \cite{HEWITT2020b}.  As boundary conditions for the salt lake scenario we take:
\begin{gather*}
S = 1, \ q_z = E \ \mathrm{on} \ z = 0, \\
S \rightarrow 0, \  q_z \rightarrow E \ \mathrm{as} \ z\rightarrow -\infty. 
\end{gather*}

In order to non-dimensionalize the governing equations, one needs to specify a characteristic length scale $L$ and velocity scale $\mathcal{V}$, with a natural time scale following as $T=\varphi L/\mathcal{V}$.  Taking advantage of how the evaporation rate sets the average vertical fluid flux everywhere, we use $\mathcal{V}=E$.  Then, in the absence of any geometric length scale in the problem (\textit{e.g.} the layer thickness often used for two-sided convection \cite{HEWITT2020b}), we set $L = \varphi D/E$ as the distance over which advective and diffusive effects will balance, for fluids moving at the characteristic speed $E$.  Rescaling Eqs.~(\ref{dim1})--(\ref{dim3}) then leads to
\begin{align}
	\bm{\nabla} \boldsymbol{\cdot} \mathbf{U} &= 0\label{eq:continuity},\\
	\frac{\partial S}{\partial \tau} &= \nabla^2 S - \mathbf{U} \boldsymbol{\cdot} \nabla S\label{eq:advection_diffusion}, \\
	\mathbf{U} &= - \nabla P - \text{Ra} S\hat{\bm{Z}}\label{eq:darcy_flow},
    \end{align}
with a rescaled velocity \mbox{$\mathbf{U}=\mathbf{q}/E$}, depth \mbox{$Z = zE/\varphi D$}, time \mbox{$\tau = tE^2/\varphi^2D$} and pressure \mbox{$P = \frac{\kappa}{\varphi \mu D}(p+\rho_0 g z)$}.
This system of equations is governed by a single dimensionless group, the Rayleigh number:
\begin{align}
\text{Ra} = \frac{\kappa \Delta \rho g}{ \mu E}\;.\label{eq:rayleigh}
\end{align}
Here, the Rayleigh number also gives the ratio of the speed at which a large heavy plume will fall under its own weight, $\mathcal{V}_\mathrm{B} = \kappa \Delta \rho g / \mu$, to the upward drift due to surface evaporation: $\text{Ra} = \mathcal{V}_\mathrm{B}/E$.  Thus, it is frequently used to characterize the vigor of porous medium convection with a through-flow \cite{WOODING1960a,HOMSY1976,WOODING1997b,VANDUJIN2002,LASSER2021}.

A more detailed derivation of this model, along with discussion of the assumptions behind it, a linear stability analysis of the onset of convection and a two-dimensional numerical investigation into its statistically steady state, was recently published by some of us \cite{LASSER2021}.  Our aim here is to show how solutal convection beneath a salt lake will couple to the growth of patterns in the crust and to make predictions which can be compared to field measurements.  As such, after a discussion of how the emergent dynamics of this model will appear in geophysically relevant conditions, we will develop predictions of the salt flux into the crust, which can give rise to crust patterns.

\subsection{Convective dynamics and scale selection\label{subsec:scale_selection}}
The model above is designed to match the expected dynamics of groundwater in a dry salt lake with at least a thin salt crust developing on top of a fluid-saturated soil.  For the one-dimensional case, where salinity depends only on depth, there is a unique stationary solution to Eqs.~(\ref{eq:continuity})--(\ref{eq:darcy_flow}), namely $S = e^{Z} = \exp(zE/\varphi D)$, which represents a salt-rich layer of groundwater lying just below the surface and a balance between advective and diffusive transport of salt \cite{WOODING1960a}.  In the absence of any horizontal flows other plausible initial conditions, such as an initially uniform lake (\textit{i.e.} $S=0$ everywhere), will relax towards the stationary solution on the characteristic timescale $T$ \cite{WOODING1997b,LASSER2021}.  Similarly, an initially crust-free lake can be described at first by an unsaturated boundary at $Z=0$, with no flux into the surface until the salinity there reaches saturation: this scenario will also develop a boundary layer that reaches saturation, and starts to precipitate, over the same timescale \cite{WOODING1997b}. {For representative parameter values from Owens Lake, $T$ is about 6 months, consistent with the known timescale of crust development~\cite{NIELD2015}, and $L = 15\pm 8$ cm.} 

For a wide variety of initial conditions, therefore, a dry salt lake will naturally develop a heavy salt-rich boundary layer of fluid near its surface.  Above some critical Rayleigh number Ra$_\mathrm{c}$ this layer will be unstable to convection \cite{WOODING1960a,HOMSY1976,WOODING1997b,VANDUJIN2002,LASSER2021}.   Linear stability analysis shows that for constant, uniform evaporation at the surface, the stationary solution has a finite-wavelength instability above Ra$_\mathrm{c} \simeq 14.35$ that leads to the growth of convective plumes \cite{WOODING1960a,HOMSY1976,LASSER2021}.  At Ra$_\mathrm{c}$ the critical wavenumber $a_\mathrm{c} \simeq 0.76$ {for conditions at Owens Lake, corresponding to a wavelength of 1.3$\,$m. Intriguingly, this characteristic scale of the convective instability is of the same order of magnitude as the patterns typically seen in salt crusts.}

This instability is also robust to different assumptions about the initial and boundary conditions of the lake.  For example, an initially uniform lake (where $S=0$ at $t=0$) has the same Ra$_\mathrm{c}$ as the stationary state, along with a very similar spectrum of unstable modes \cite{LASSER2021}. Alternatively, as would be appropriate for a salt lake with a thin ponding of surface brine, the case of a constant pressure boundary condition at $Z = 0$ is unstable to convection above Ra$_\mathrm{c} \simeq 6.95$, with $a_\mathrm{c}\simeq 0.43$.   

To predict the long-time dynamics of convection pattern in a dry salt lake, we simulated Eqs.~(\ref{eq:continuity})--(\ref{eq:darcy_flow}) on rectangular domains (2D) or boxes (3D), where $S = U_Z = 1$ at the top boundary, $S = 0$ and $U_Z = 1$ at the bottom, and with periodic boundary conditions along the vertical walls. We conducted simulations from $\mathrm{Ra}=20$ to 1000. As initial conditions, we started from the stationary state for a bounded domain, $S = (e^{Z}-e^{-h})/(1-e^{-h})$, with small random perturbations applied to the salinity distribution at all grid points.  {Most simulation domains had a horizontal extent of 12$\pi$ (for 2D) or 24$\pi$ (for 3D) and depth of $h = 10$.  The domain depth introduces an additional length scale, and thus dimensionless group, into the numerical problem.  To confirm that the results are in a limit where they are essentially independent of $h$, we repeated selected simulations for domains with $h=20$.}   Our numerical methods are summarized in Appendix~\ref{app:simulation}. 

\begin{figure*}
\centering
\includegraphics[width=1\textwidth]{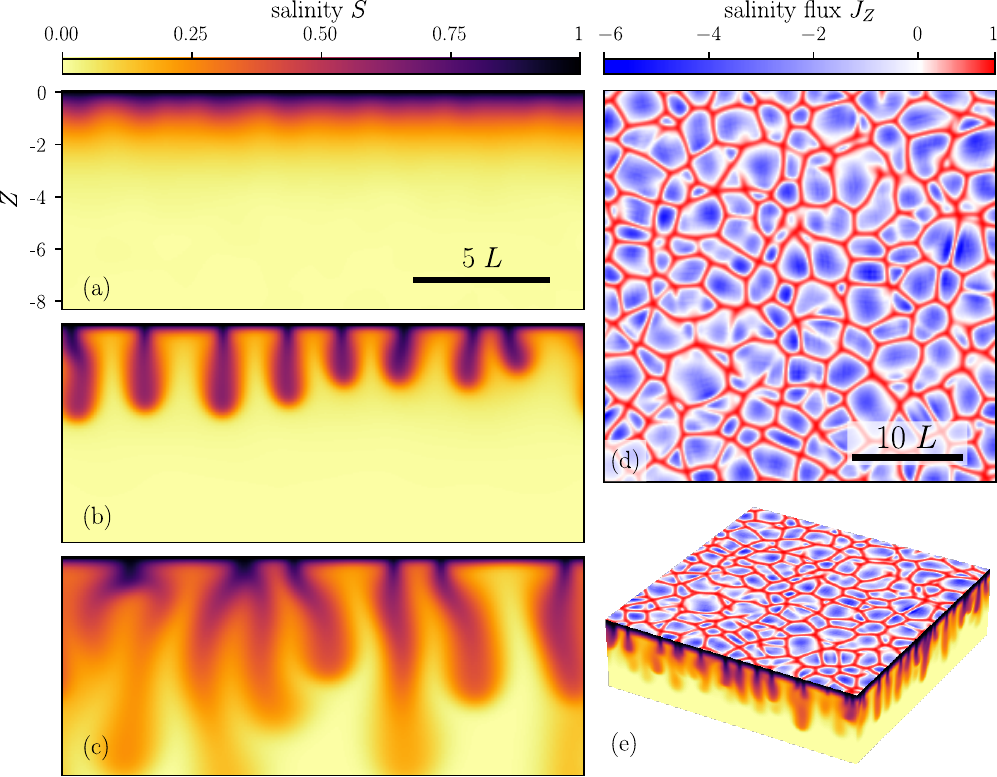}
\caption{\label{fig:fig3} The dynamics of porous media convection simulated below a salt lake are shown for $\mathrm{Ra}=100$.  For two-dimensional simulations, the development of high-salinity ($S$) downwelling plumes is shown for times \textbf{(a)} $\tau=0.1$, \textbf{(b)} $\tau=0.3$ and \textbf{(c)} $\tau=1.0$. Supplemental movie S2 shows the full dynamics of the simulation. Panel \textbf{(d)} shows a snapshot of three-dimensional salinity flux into the surface ($Z=0$) for $\tau = 1.0$. Panel \textbf{(e)} shows the same snapshot, now including a view of the downwellings of high salinity along the vertical faces of the domain.  Here, the downwellings arrange into narrow sheets, forming a network around more diffuse upwelling regions.  Supplemental movie S3 shows the full dynamics of the three-dimensional simulation.}
\end{figure*}

A finite-wavelength instability sets in for all simulations and rapidly grows beyond the linear regime.  As shown for the 2D case in Fig.~\ref{fig:fig3}(a--c), perturbations around the most unstable mode develop into a set of narrow plumes of high salinity, which fall from the upper surface.  In 3D the situation is similar, as in Fig.~\ref{fig:fig3}(d,e), with the convection near the surface instead organizing into a closed polygonal network of downwelling sheets, which surround patches of upwelling fluid.  At lower depths the sheets break up into isolated plumes.  Both the growth rate of the linear instability and the speed of the mature downwelling features scale with $\epsilon = (\mathrm{Ra}-\mathrm{Ra_c})/\mathrm{Ra_c}$, which can be understood as $\mathrm{Ra} \sim \mathcal{V}_\mathrm{B}$.  As such,  we make use of a rescaled time, $\widehat{\tau} = \tau\epsilon$, in order to compare features at a similar stage of development.      

As they grow, the initial downwelling features strip the boundary layer of dissolved salt, leaving it depleted of the buoyancy forces that initiated the convection. The plumes also tend to merge together into larger structures.  {This coarsening process causes the pattern length-scale to increase over time; for further details, see Ref.~\cite{LASSER2021}}.  However, as the distance between downwelling areas grows, they become less efficient at draining salinity from the boundary layer, which can start to thicken again.  This allows for the growth of additional instabilities.  In 2D these take the form of small plumes that appear between, and are attracted to, the larger established downwellings.  In 3D, although new point-like plumes do also occur, the initiation of linear features is more typical. As shown in Fig.~\ref{fig:fig3}(d,e), these are weaker downwelling sheets that connect two sides of polygonal convection cells.  As they grow, they also move towards the boundaries of the more established, larger convection cells, eventually merging with an existing edge.  {These transient `microplume' features become more prevalent at higher Ra, overlaid on the more stable `megaplumes' of convection cells~\cite{Paoli2022}, which extend much deeper than the boundary layer. Over long enough times a statistically steady state develops~\cite{LASSER2021} where the rates of plume formation and merging balance, and a well-defined length-scale emerges}.  

{These dynamics can be compared with analogous results from two-sided convection without throughflow~\cite{HEWITT2012,FU2013,HEWITT2014,Paoli2022}, which has become an important model of carbon geosequestration.  For this, some caution is required, as the characteristic scales and Rayleigh number are defined differently in the two cases.  Comparing the two systems, the patterns near the walls are remarkably similar, with transient ribs or sheets appearing between larger, more stable and regular polygonal cells~\cite{FU2013,HEWITT2014,Paoli2022}.   Interestingly, however, the wavelength selection process is quite different.  For two-sided convection the dominant circulation cells, sampled at the midplane, scale as $a\sim\mathrm{Ra}^{0.5}$.  Near the walls the density of the smaller, transient features increases more rapidly with Ra, with a dependence thought to approach $a \sim \mathrm{Ra}$ at high enough Ra~\cite{FU2013,HEWITT2020b,Paoli2022}.  However, averaging out these fast-moving features over time, or through spatial filtering, shows a near-surface pattern that closely matches the internal convection cells~\cite{Paoli2022}.} 

\begin{figure}
\centering
\includegraphics[width=0.47\textwidth]{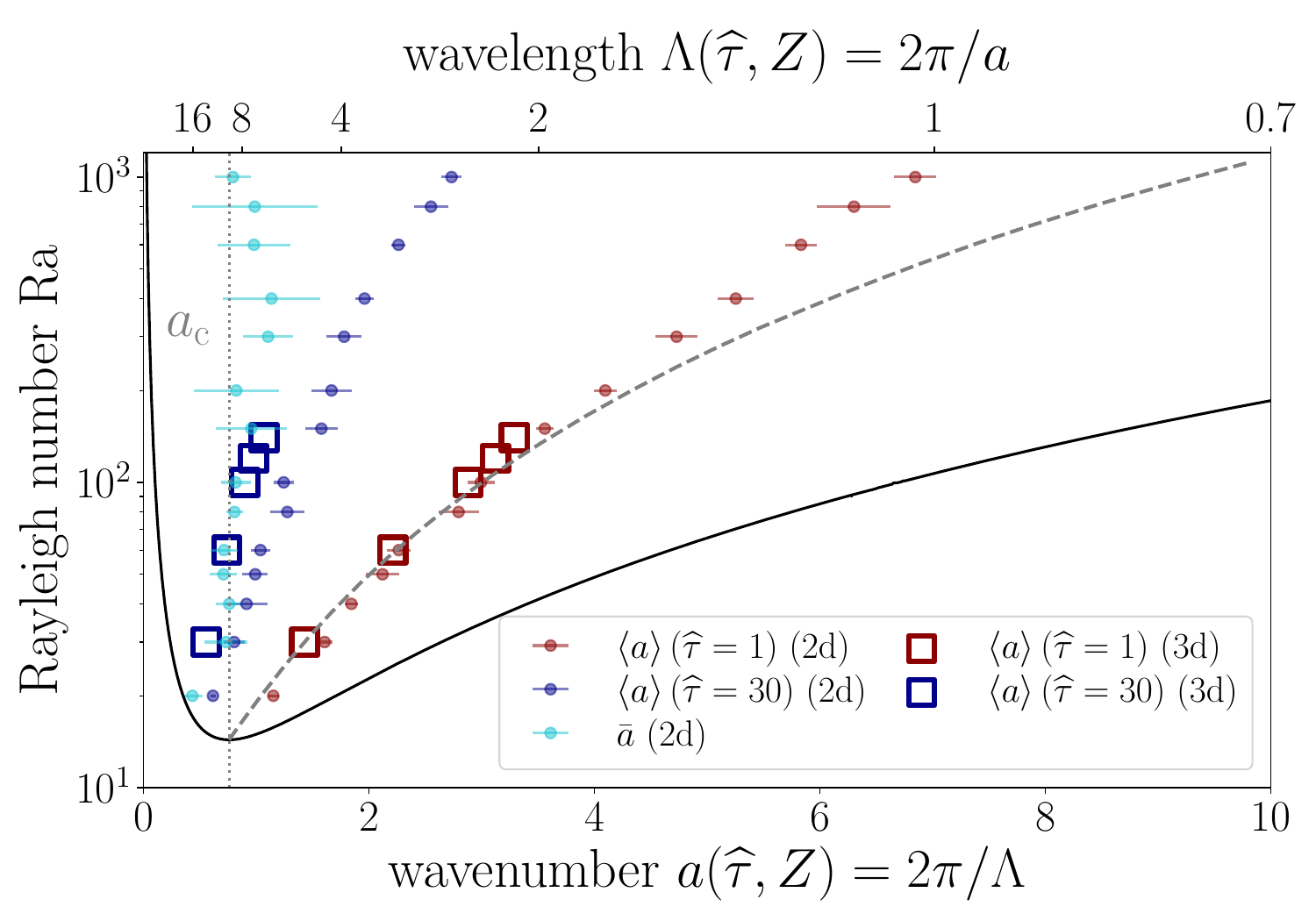}
\caption{\label{fig:fig4} Pattern wavenumbers $a$, as measured in two-dimensional (circles) and three-dimensional (squares) simulations for different $\mathrm{Ra}$. Average wavenumbers $\left<a\right>$ are measured at two times: once early into the growth of the initial instability, at $\widehat{\tau} = 1$ (red), and later at $\widehat{\tau}=30$ (blue), after the pattern has had time to develop towards a statistically steady state. {We also report the wavenumber of the dominant mode $\bar{a}$ (turquoise), at late times.} For the 2D simulations, the marker positions and error bars represent means and standard deviations as measured in ensembles of 6--10 realizations each.}
\end{figure}

For our one-sided salt lake model, quantitative details of the scale selection process are presented in Fig.~\ref{fig:fig4}, which also includes the neutral stability curve and most unstable mode of convection, as derived in Ref.~\cite{LASSER2021}. Simulations with $\mathrm{Ra} > \mathrm{Ra}_\mathrm{c}$ are unstable to convective overturning, which becomes more vigorous with increasing Ra.   {To begin with, pattern wavenumbers were calculated as in \cite{HEWITT2014}, as the first moments of the radially-averaged power spectra of the salinity distributions at a depth of $Z \approx 1$.}  The initial instability was characterized at the rescaled time $\widehat{\tau} = 1$. At these early times the spacing of downwellings in both the 2D and 3D simulations (red markers) agrees well with the most unstable mode of the linear stability analysis. When measured at $\widehat{\tau}= 30$, however, many downwelling plumes have merged, leading to smaller wavenumbers (blue markers).  
{Confirming that the system is in a statistically steady state by this time, simulations run to $\widehat{\tau}= 60$ show no measurable systematic drift in $k$ (values match with $\widehat{\tau}= 30$ to $\Delta k = -0.004 \pm 0.01$, mean $\pm$ standard error).  Similarly, additional simulations with $h=20$ gave comparable results (values match with $h = 10$ to $\Delta k = -0.04 \pm 0.04$).}

{In the steady state, there is a gradual increase in the average wavenumber with Ra, as is the case also for 2-sided convection, although this trend is systematically weaker than the scalings evident in that system~\cite{FU2013,HEWITT2014,Paoli2022}.  This increase appears to be driven by the proliferation of small, transient downwelling microplumes as Ra increases.  These intermittently appear, then are swept into the more established large-scale convective features (see supplemental movies S2, S3).  As the surface crust is built up over time, we would expect it to reflect the more stable features, \textit{i.e.} megaplumes. To isolate the size of these features, we calculated the dominant wavenumber, $\bar{a}$, associated with the maximum of the power spectrum.  As shown in Fig.~\ref{fig:fig4} (turquoise markers), this predicts a pattern of polygonal cells, bounded by salt-rich downwellings, and a characteristic wavelength that consistently settles to values near to $\Lambda = 2\pi/a_c$, or about 1.3$\,$m for the conditions typical of Owens Lake.}


\subsection{Salt flux and ridge growth
\label{subsec:salt_flux_and_crust_growth}}

At the surface of the soil water will evaporate and leave behind its dissolved burden of salt.  In our model of this process the rate at which salt is added to the surface depends on its advection along with the water, but is also influenced by diffusion down concentration gradients away from the crust.  In other words, the mass balance of Eq.~(\ref{eq:advection_diffusion}) can be written in terms of a salinity flux
\begin{equation}
    \bm{J}_S = \bm{U}S - \bm{\nabla} S \label{J_s}
\end{equation}
where $\partial S/\partial \tau = -\bm{\nabla \cdot J}_S$.  At the surface, $Z = 0$, the upwards water flux matches the evaporation rate, so $U_Z = 1$.  The presence of a salt crust also sets $S = 1$ there.   As such, the salinity flux into the crust is given by 
\begin{equation}
    J_Z = 1 - \left.\frac{\partial S}{\partial Z}
    \label{S_flux}\right|_{Z=0}.
\end{equation}
This flux vanishes, $J_Z = 0$, in the stationary solution $S = \exp(Z)$, although this scenario will still leave a constant upward flux of \textit{salt} into the crust at whatever concentration is supplied by the reservoir, as we will describe shortly.  Variations in the near-surface salinity gradient will then modulate crust growth around this reference case.  Where vertical gradients are strong crust growth will be suppressed ($J_Z<0$), whereas over weaker gradients crust growth will be enhanced ($J_Z>0$).  
\begin{figure*}
\centering
\includegraphics[width=\textwidth]{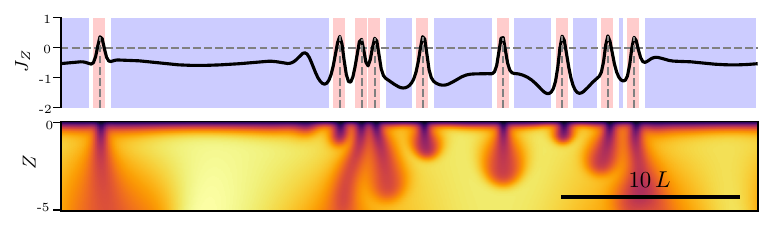}
\caption{\label{fig:fig5} Salinity flux ($J_Z = 1-\partial S/\partial Z$) into the modeled surface crust. The simulation snapshot is for $\mathrm{Ra} = 100$ at time $\widehat{\tau} = 150$.  Areas predicted to experience enhanced ($J_Z>0$) or suppressed ($J_Z<0$) salt flux are indicated as shaded red and blue areas, respectively. Peaks, defined as local maxima with a flux value above zero, are indicated as dashed lines and lie directly above downwelling plumes of high salinity; the lower panel gives the salinity $S$ in the domain at the same time.}
\end{figure*}

To demonstrate this modulation quantitatively, we calculated the surface salinity flux according to Eq.~(\ref{S_flux}) in simulations at Ra = 100, with results for 2D shown in Fig.~\ref{fig:fig5}, and for 3D shown in Fig.~\ref{fig:fig3}(d, e). Supplemental movies S2 and S3 show the full dynamics of the two- and three-dimensional systems, respectively. Above downwellings plumes (2D) or sheets (3D), the salinity flux into the surface is positive and shows marked peaks, whereas between downwellings the salinity flux into the surface is negative, and roughly constant.  As sketched in Fig.~\ref{fig:fig2}, this predicts the development of narrow regions of faster salt precipitation above any convective downwellings, which we argue gives rise to ridges there.

As visible in the movies S2 and S3, the positions of downwelling plumes slowly fluctuate as the simulations progress, as does the pattern of flux into the crust. The motion of the more well-established plumes is gradual, especially in 3D, but the presence of large ridges at some field sites suggests that feedback mechanisms may be needed to help stabilize the patterns over long periods of time. Such feedback could act through a spatial modulation of the evaporation rate $E$ based on salt coverage, which then acts to pin the position of the downwellings in space. Reference~\cite{LASSER2021} gives some discussion on how such a feedback mechanism could work.

We also note that the dynamics in a sufficiently deep lake should approach the case where the average surface salinity flux into the crust vanishes.  This follows from the divergence theorem.  For a deep lake the descending plumes will mix diffusively with upwelling fluid over a timescale $\sim \Lambda^2$, for plume spacing $\Lambda$.   Since the plumes start their fall from the surface at a characteristic velocity $\mathcal{V}_\mathrm{B}\sim\mathrm{Ra}$~\cite{LASSER2021}, this mixing will occur over depths of order $\mathrm{Ra}\Lambda^2$.  As they mix and lose buoyancy the plumes will slow, setting a maximum depth that they can penetrate, of the same order.  Below this depth the average upwards salinity flux of the flows from the reservoir must approach zero.  Since in a statistically steady state the total salinity in the convecting part of the domain is constant, or fluctuating around a well-defined average value, the salinity flux into the surface must also approach zero.  Probing this limit accurately, however, requires domain sizes well beyond the scope of our current simulations.

{To make predictions of the crust dynamics at any particular field site, the salinity flux $J_Z$ needs to be converted into estimates of the real salt flux into the crust.}  Assuming minimal volume change due to mixing, the density of the groundwater can be written as $\rho = \rho_w + \beta c$, where $\rho_w$ is the density of pure water, $c$ is the mass concentration of dissolved salts in the solution, and $\beta$ is a coefficient of expansion.  Analogous to the salinity flux in Eq.~(\ref{J_s}) the mass flux of salt, in dimensional variables, is 
\begin{align}
    \bm{J}_c &= \bm{q}c - D\varphi\bm{\nabla} c, \label{eq:salt_fluxA} \\ & = \bm{q} c_0 + \frac{\Delta\rho}{\beta}E\bm{J}_S,
\end{align}
where the second equation follows by changing variables using $\rho = \rho_0 +S\Delta \rho$.  As can be seen, there is an offset between the two fluxes--representing the fact that the groundwater feeding into the system from a distant source ($S=0$) contains some initial level of dissolved salt, $c_0$.   In dimensional terms, the predicted salt mass flux into the crust is now
\begin{align}
    {J}_{c,z} &= Ec_{\mathrm{sat}} - D\varphi\left.\frac{\partial c}{\partial z} \right|_{z=0}. \label{eq:salt_flux}
\end{align}
We will apply this to Owens Lake in Section \ref{subsec:evidence_surface_salt_flux}.

\subsection{Model summary \& predictions \label{subsec:model_predictions_and_summary}}
We have developed here a model of porous medium convection beneath a salt lake, which can accurately predict much of the known phenomena of the crust patterns in dry salt lakes, as summarized in Section \ref{sec:phenomenology_of_salt_polygons}.  The driving mechanism of salinity-driven convection and its associated governing equations are drawn from existing models of salt lake dynamics \cite{WOODING1997b,WOODING2007,LASSER2021} and observations of convection beneath tidal sabkhas \cite{STEVENS2009,VANDAM2009}, but we extend these to consider interactions with the surface crust. Evaporation at the surface causes a vertical salinity gradient to emerge, which leads to salt precipitation and crust growth. If a critical Rayleigh number is surpassed, convective cells emerge and modulate the salinity flux into the surface, with faster precipitation expected over downwellings and slower crust growth over upwellings.   
We will provide field observations in Section~\ref{sec:evidence} to quantify these predictions, and show that they are consistent with average growth rates {on the order of millimeters per month}, and about a two-fold faster deposition of salt at downwellings, relative to upwelling regions.  Observing the pattern structure in 3D, we found that over sufficient time the downwelling features arrange into narrow sheets that delimit a closed polygonal pattern of convection cells.  The scale of these polygons, in a steady state, does not have any strong dependence on the Rayleigh number. {In particular, from Ra = 20 to 1000 we showed that the average wavenumber of the near-surface convection pattern changes by at most a factor of four; the dominant wavenumber, which reflects the structure of the more stable, deeper convection cells, does not evolve significantly over this range and remains consistent with the critical wavenumber, $a_c = 0.76$.} 
This predicts that the pattern wavelength will be relatively insensitive to soil composition (which determines \textit{e.g.} permeability), groundwater chemistry (\textit{e.g.} sulphate rich, or pure halite, affecting $\Delta \rho$ and to a lesser extent $D$) and robust to fluctuations in environmental driving parameters.

\section{Evidence for convection as driving mechanism}\label{sec:evidence}
In Section~\ref{sec:phenomenology_of_salt_polygons} we outlined a set of criteria against which to test mechanisms of pattern formation in salt playa.  Briefly, any mechanism should be active in salt lake settings, modulate salt transport by directing more salt towards ridges, account for the timescales of crust growth, explain the consistent wavelength of closed polygonal shapes in the crust and be robust enough to explain all this across the variety of conditions associated with patterned crusts.  Here, we will explore these criteria in turn, focusing on empirical evidence from field work, supplemented by analogue experiments.  Details of field data collection are given in Appendix~\ref{app:field}, data processing in Appendices~\ref{app:grain_size}--\ref{app:salt_flux} and experimental methods in Appendix~\ref{app:experiments}.  The results point towards convective overturning as a suitable driving mechanism of salt crust patterns.  

\subsection{Convective instability \label{subsec:evidence_onset_of_convection}}
A driving mechanism should be specific to the geophysical conditions encountered in salt lakes.  As our model is based on boundary conditions and governing equations of porous media transport that are appropriate for playa and sabkhas (\textit{e.g.} \cite{WOODING1997b,BOUFADEL1999,VANDUJIN2002,WOODING2007,LASSER2021}), we aim to prove here that typical dry salt lakes have conditions that would make them unstable to the onset of convective motion.  To this end, we visited and characterized field sites at Owens Lake (CA, USA), Badwater Basin (CA, USA) and Sua Pan (Botswana), to show that they can be described by a Rayleigh number above the critical value of $\mathrm{Ra}_\mathrm{c} \simeq 14.35$ \cite{HOMSY1976,VANDUJIN2002,LASSER2021}. All sites show well-developed polygonal patterns, and details of the field methodology are given in Appendix~\ref{app:field}.  These results are supplemented by experiments that visualize buoyancy-driven convective dynamics in conditions similar to those of real salt lakes (see Supplemental movie~S4).

To determine the Rayleigh number appropriate for each field site, we measured the relevant parameters of Eq.~(\ref{eq:rayleigh}).  From grain-size distributions (data deposited at~\cite{GRAINSIZES}, analysis in Appendix~\ref{app:grain_size}) we calculated $d_s$, the Sauter diameter~\cite{SAUTER1928}, of near-surface soil samples.  The results, from $4.3\pm 0.6\,\mu$m to $138\pm23\,\mu$m, represent a silt to fine sand. A high soil porosity, $\varphi=0.70\pm0.02$, was previously measured at Owens Lake~\cite{TYLER1997}. From these, the relative permeability was estimated using the empirical relationship \mbox{$\kappa = 0.11\,\varphi^{5.6}\,d_s^2\,$}, which fits a broad set of experimental and simulation data~\cite{GARCIA2009}. Across all sites \mbox{$\kappa=(3\pm 2) \times 10^{-13}\,$m$^2$} to $(2.7 \pm 1.2)\times 10^{-10}\,$m$^2$. At Owens Lake we also directly measured density differences of \mbox{$\Delta\rho=210\pm10\,$kg\,m$^{-3}$} between pore water samples taken from close to the surface and at approximately $1\,$m depth (data at~\cite{CHEMICAL}). {Annual average evaporation rates of groundwater are taken as $E=0.4\pm 0.1\,$mm/day~\cite{GROENEVELD2010,TYLER1997} for Owens Lake, $0.3\pm 0.1\,$mm/day~\cite{DEMEO2003} for Badwater Basin, and $E=0.7\pm0.5\,$mm/day for Sua pan~\cite{BRUNNER2004,NIELD2016}, 
with further details reviewed in Appendix~\ref{app:evap}. These low levels of groundwater evaporation are characteristic of salt pan environments, and similar rates are seen in active playa of the Atacama desert ($E=0.5\pm 0.1\,$mm/day~\cite{KAMPF2005}) and sabkha near Abu Dhabi ($E=0.2\pm 0.05\,$mm/day~\cite{WOOD2002}).} Finally, we assume the dynamic viscosity of the groundwater to be a constant $\mu = 10^{-3}\,$Pa$\,$s. A detailed description of the data sets from Owens Lake and Badwater Basin is in Ref.~\cite{lasser2020surface}.

From these observations we calculated $\mathrm{Ra}$ at twenty-one sites around Owens Lake, five in Badwater Basin and seven at Sua Pan. The median values at these three regions were $\mathrm{Ra}=3700$, $36000$ and $420$, respectively. Values for all 33 sample locations were between $\mathrm{Ra}=120\pm 50$ and $(1.2\pm 0.5)\times 10^5$, well above Ra$_\mathrm{c}$. The Sauter diameters, permeabilities and Rayleigh numbers for all sites are given in the Appendix, Table~\ref{tab:tab4}. The conditions throughout these patterned salt playa are, therefore, suitable to expect a convective overturning of their groundwater, with plumes of high salinity sinking downwards from the surface.  

It is interesting to note that convective plumes of salt-rich water have also been observed by electrical resistivity measurements after a heavy rainfall on salt crusts near Abu Dhabi~\cite{VANDAM2009} and beneath wind-tidal flats in Texas~\cite{STEVENS2009}.  While these are a slightly different phenomenon, involving the sudden addition of salt-rich brine to the surface as the rain dissolves salt, they demonstrate convection in similar geophysical conditions to salt lakes.  The Rayleigh numbers calculated for these two cases are up to about 40,000 and 90, respectively~\cite{VANDAM2009,STEVENS2009}.

We complemented these observations with analogue experiments in Hele-Shaw cells to demonstrate buoyancy-driven convection in porous media under conditions comparable to the field.  These experiments are inspired by Refs.~\cite{WOODING1997a,SLIM2013,THOMAS2018} but instead of using a narrow and empty gap between the cell walls to create the porous medium, we used a 0.8 cm thick cell filled with glass beads.  We applied a strong surface evaporation, driven by heat lamps and ventilation, and the base of the cells were connected to fluid reservoirs with a fixed salt concentration of 50 kg\,m$^{-3}$.  Further details of the experimental methods are given in Appendix~\ref{app:experiments} and an example time-lapse movie of an experiment in progress is given as supplemental movie S4.  To observe whether a convective instability arises for different conditions, we systematically varied the grain sizes of the glass beads.  This modifies the relative permeability of the system and allowed us to change the Rayleigh number of the experiments.  Convection was seen in all experiments above Ra$_\mathrm{c}$, but not in the finest-grained case, where $\mathrm{Ra}\ll\mathrm{Ra_c}$.  These results are consistent with those of Wooding \textit{et al.} \cite{WOODING1997a}, who experimentally confirmed theoretical predictions of the onset of convection for the slightly different case of a salt lake with surface ponding, or constant pressure boundary conditions.  Finally, we note that the range over which the experiments show convection, between $\mathrm{Ra} = 20\pm10 $ and $(1.3\pm 0.3) \times 10^3$, also coincides with much of the range of Rayleigh numbers we estimated from the field sites.

\subsection{Mapping salt heterogeneity\label{subsec:evidence_salinity_distribution_heterogeneities}} 
If salt polygon growth is driven by convective dynamics happening beneath the surface, then horizontal differences in salt concentration should be detectable in soil and pore fluid under typical field conditions, and also in laboratory-based analogue experiments.   Salt variations of this nature have been seen in sabkhas after rain \cite{VANDAM2009,STEVENS2009}.  We provide evidence here of salt-rich plumes beneath the crust of Owens Lake, and show that the plumes are correlated with the positions of the ridges in the crust.  

\begin{figure}
\centering
\includegraphics[width=0.47\textwidth]{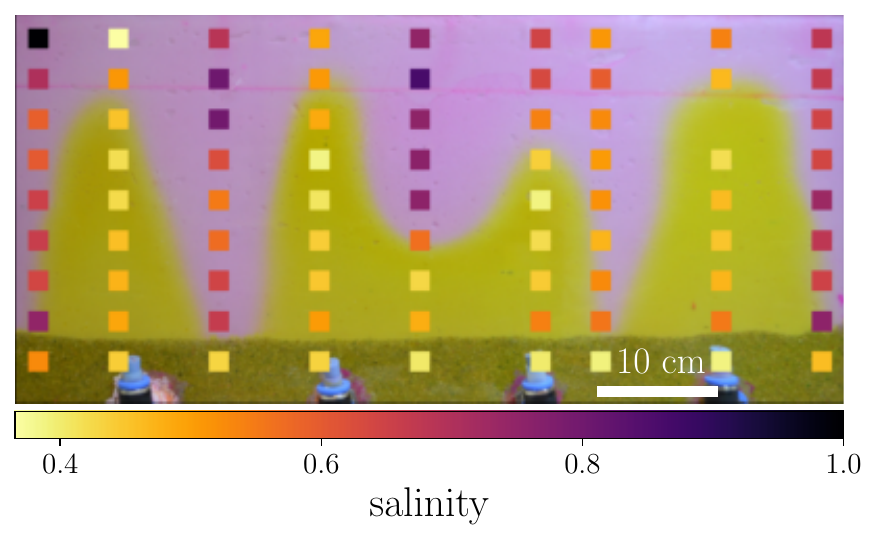}
\caption{\label{fig:fig6} Convective plumes in an experimental Hele-Shaw cell are highlighted by dye (the brighter upwelling fluid results from dying the reservoir, well after convection has set in).  The motion is coupled to the salinity, which was measured at the colored squares by destructive sampling.}
\end{figure}

First, to demonstrate the coupling of salt concentration and convective motion under controlled conditions, we sampled one of the Hele-Shaw experiments introduced in Section ~\ref{subsec:evidence_onset_of_convection}.  For this we dissected an experiment that was undergoing convection, and which had been evolving under constant conditions for two months.  Samples were extracted from locations along the downwelling and upwelling plumes, as indicated by the motion of dye added into the reservoir fluid a few days before sampling.  As shown in Fig.~\ref{fig:fig6}, the fluid flow in the analogue experiments is clearly driven by, and coupled to, variations in salinity.  Pore water samples taken from the upwelling regions (dyed yellow in the figure) show systematically lower salinity than those taken from downwellings.    

\begin{figure*}
\centering
\includegraphics[width=\textwidth]{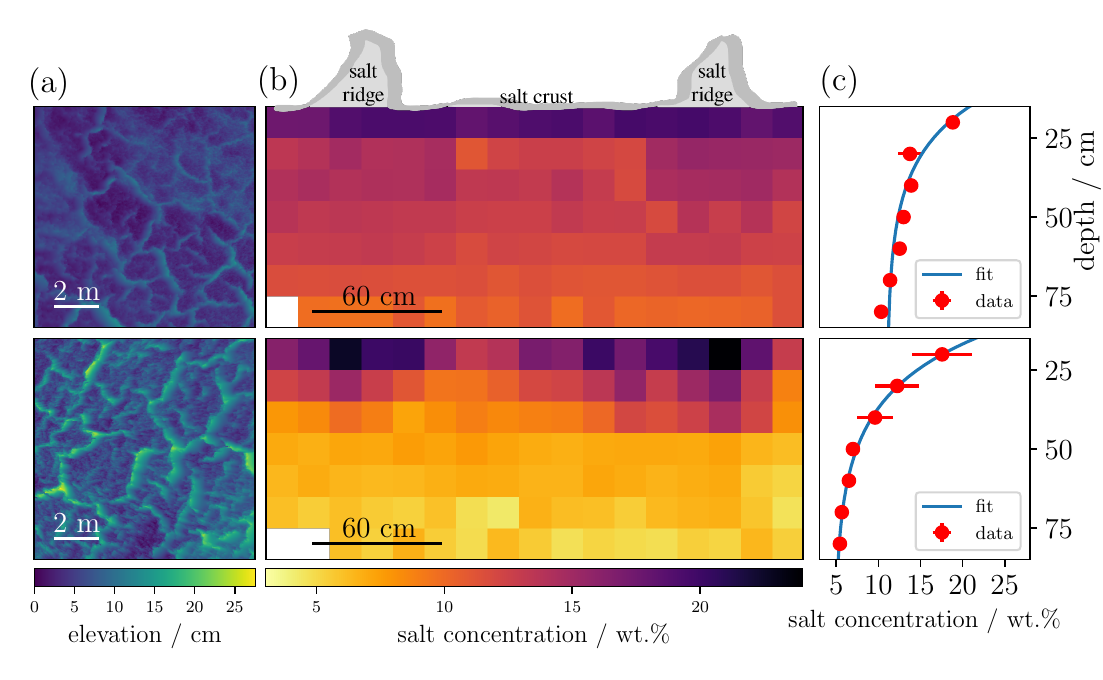}
\caption{\label{fig:fig7} Surface height maps and subsurface salinity profiles for two sites (site 1, top; site 2, bottom) at Owens Lake. \textbf{(a)} TLS scans give the elevation of ridges on the surface crust. \textbf{(b)} Cross-sections of polygons at Owens Lake show the variation of salt concentration with depth and relative to the ridge positions. Each square corresponds to the salt concentration in the pore fluid of an individual field sample. \textbf{(c)} Exponential fits (see Appendix~\ref{app:salt_flux}) to the changing salt concentration with depth. Red dots represent horizontally averaged salt concentrations from data in \textbf{(b)}, and error bars show the standard deviation.}
\end{figure*}

Next, from the field we collected samples of wet soil from fresh trenches dug at two unmanaged sites at Owens Lake.  Surveys of the surface relief were made before sampling, using a terrestrial laser scanner (TLS, see Appendix~\ref{app:tls} and Ref. \cite{NIELD2013} for methodology; data deposited at~\cite{TLS}).  These show the presence of salt polygons of about $2\,$m in size delimited by high ridges, as in Fig.~\ref{fig:fig7}(a). In each case we sampled soil along a grid pattern in a cross-section below a polygon, including beneath two bounding ridges. Analysis of the salt concentration of the samples with respect to pore water content shows direct evidence of plumes of high-salinity fluid  below the salt ridges (Fig.~\ref{fig:fig7}(b), methods in Appendix~\ref{app:salinity} and data deposited at~\cite{CONCENTRATION}). Specifically, we tested whether the distribution of salt concentrations in an area below each ridge was different to that below the flat pan of the polygon; testing this hypothesis (two-sample KS test), shows that the distributions below ridges and flat crust are statistically distinct ($p < 0.02$), at both sites.

The measurements of the salt concentration in the pore water also shows an exponential decay in salinity with depth (Fig.~\ref{fig:fig7}(c)), consistent with a salt-rich boundary layer that is heavy enough to drive convection in a porous medium.  As evidence for this, we recovered representative values for the boundary layer thickness from exponential fits to the horizontally-averaged salt concentrations at both trench sites (see Appendix~\ref{app:salt_flux}).  Interestingly, the observed values of $13.5\pm 5.3\,$cm and $17.7\pm1.5\,$cm are comparable to the natural length scale of $L=\varphi D/E = 15.1\pm8.0\;$cm estimated for Owens Lake.    

Thus, not only does direct field sampling of groundwater beneath a patterned salt crust show both horizontal and vertical variations in salt concentration, which support the hypothesis that the system is unstable and convecting, but it also demonstrates that the plumes of high salinity are co-localized with the surface ridges.

\subsection{Surface salt flux in the field
\label{subsec:evidence_surface_salt_flux}}
A driving mechanism for pattern formation should be able to spatially modulate salt transport to the soil surface, leading to salt ridge growth on a scale of months. We demonstrate this growth at Owens Lake through a time-lapse video (Supplemental movie S1) which shows the surface crust dynamics over a period of approximately four months in spring 2018. {At Sua Pan, crust dynamics have been measured by TLS scanning of several sites over the course of several months~\cite{NIELD2015}. Active crusts there showed average growth rates of 0.7 to 1.5 mm/month.  The fastest short-term growth rates of ridge features, briefly reaching up to 10 mm/month, were associated with ridge thrusting.} In Section~\ref{subsec:salt_flux_and_crust_growth} we explored the corresponding predictions of our model. Specifically, Eq.~(\ref{eq:salt_flux}) describes the expected salt flux into the surface, $J_{c,z}$, depending on the salt concentration gradient there. Here, we apply the framework of this model to the maps of the subsurface salt concentration measured at Owens Lake (Fig.~\ref{fig:fig7}). In particular, we will show that the observed variations are sufficient to drive {average crust growth rates on the order of millimeters per month along with} a preferential growth of the ridges, over the rest of the crust.

To calculate the expected salt flux into the surface, we estimated the near-surface gradient in salt concentration from the field data from our trench sites. Dividing this data between regions beneath the ridges and centers of each polygon, we used an exponential fit to recover the effective thickness of the salt-rich boundary layer $L'$, the background salt concentration $c_\mathrm{bkg}$ and the salt concentration at the surface $c_\mathrm{sat}$, for each such region.  For this, the measured weight fractions of salt were re-scaled by the densities of salt solutions, to give concentrations $c$ in terms of the mass of dissolved salt per unit volume of fluid.  Further details of the fitting are given in Appendix~\ref{app:salt_flux}, including Figs.~\ref{fig:fig11} and \ref{fig:fig12}.  

Substituting the exponential form of the fit into Eq.~(\ref{eq:salt_flux}) yields a prediction for the mass flux of salt, in terms of measurable quantities, of
\begin{equation}
    J_{c, z} = E c_\mathrm{sat} -  \varphi D \frac{c_\mathrm{sat} - c_\mathrm{bkg}}{L'}\;.\label{eq:salt_flux_master_eq}
\end{equation}
Values for $L'$ and $c_\mathrm{bkg}$ below each ridge and polygon center are extracted from the fits.  For $c_\mathrm{sat}$, the fitted values are all consistent with each other, so a common value of a saturated solution, $c_\mathrm{sat}=316$ kg m$^{-3}$, was used to minimize uncertainties. As in Section~\ref{subsec:evidence_onset_of_convection} we use values for the evaporation rate, \mbox{$E = 4.6 \times 10^{-9}\,$m$\,$s$^{-1}$},  and porosity, \mbox{$\varphi = 0.7$}, taken from previous observations at Owens Lake \cite{GROENEVELD2010,TYLER1997}. For an effective diffusion constant, we use the tortuosity-corrected diffusivity \mbox{$D=1.00\times10^{-9}\,$m$^2$\,s$^{-1}$} (see Appendix~\ref{app:chemistry}).   Finally, we measured the density of salt crust samples collected from Owens Lake to be \mbox{$\rho_\mathrm{crust} = (960\pm 50)\,$kg$\,$m$^{-3}$}. This value is used to convert mass flux rates into crust growth rates.  

The resulting salt flux and growth rates for the ridges and centers of the two trench sites from Fig.~\ref{fig:fig7} are shown in Table~\ref{tab:tab1}.  The derived salt flux into the surface is about twice as high at the ridges, as compared to the centers of polygons.  In order to allow a clear statistical comparison between these rates, only independent sources of errors were propagated (\textit{i.e.} excluding uncertainties in $E$, $D$ and $\varphi$, as these will affect all rates in the same way) to give the main uncertainties reported in Table~\ref{tab:tab1}. Within these uncertainties, we see a consistent difference between the crust growth rates at polygon ridges when compared to polygon centers.  {As salt is being added faster to the ridge areas, we would expect differential growth, focused on the ridges.  Although the crust morphology can become quite complex~\cite{KRINSLEY1970,DIXON2009,NACHSHON2018}, the faster expansion of the crust in these locations can plausibly contribute to differential strains, and features like ridge thrusting or local buckling.  This is consistent with the faster dynamics seen at ridges in Sua Pan~\cite{NIELD2015} and that we report at Owens Lake (see supplemental movie S1).}

Over long times the average crust growth rate will be essentially constrained by the evaporation rate and the amount of salt in the groundwater feeding into the lake. However, it would not be surprising if these rates varied with time, e.g seasonally, or via extreme events like surface flooding. 
As we have not accounted for the excess mass of the hydrated states of any of the crust materials (especially mirabilite, present as a significant minority species of salt at Owens Lake and Sua Pan, although not noticeably present at other pans, such as Badwater Basin), the rates we report in Table~\ref{tab:tab1} also represent a lower bound on the anticipated crust growth rates.

We have shown here how the heterogeneous salinity distribution measured below salt polygons in the field infers variations in the salt flux to the surface, if the subsurface fluids are undergoing porous medium convection.  Within the framework of our convection model, the measured pattern of salt concentration near the surface of the crust imply about a twofold increase in crust growth rate at the ridges, rather than the centers of salt polygons. {} The crust growth rates {on the order of} millimeters per month are also consistent with the time scale of crust growth observed in the field, which show that noticeable changes can occur over months.

\begin{table}
    \centering
    \begin{tabular}{cccc}
         site & area & \,$J_{c, z} \;10^{-7}$ [kg$\,$m${^{-2}}\,$s$^{-1}$]\, & $r$ [mm/month] \\
         \hline
         1 & ridge 1 & $5.6 \pm 2.0\;(3.9)$ & $1.5\pm 0.6  \;(1.1)$ \\
         1 & ridge 2 & $6.8 \pm 1.2\;(3.2)$ & $1.8\pm 0.4  \;(0.9)$ \\
         1 & center & $3.3 \pm 2.5\;(4.9)$ & $0.9 \pm 0.7  \;(1.4)$ \\
         \hline
         2 & ridge 1 & $5.3 \pm 1.3\;(3.8)$ & $1.4 \pm 0.4  \;(1.1)$ \\
         2 & ridge 2 & $6.8 \pm 1.1\;(3.1)$ & $1.8 \pm 0.3  \;(0.9)$ \\
         2 & center & $2.1\pm 0.9\;(4.8)$ & $0.6 \pm 0.3 \;(1.3)$ \\
         \hline
    \end{tabular}
    \caption{Salt flux into the surface $J_{c, z}$ and crust growth rate $r$ inferred at the polygon ridges and centers for the two sites at Owens Lake depicted in Fig.~\ref{fig:fig7}. The error bounds consider independent sources of uncertainty, while the values in brackets give ranges taking into account all sources of uncertainty ($i.e.$ including lake-wide uncertainties in $E, D$ and $\varphi$).}
    \label{tab:tab1}
\end{table}

\subsection{Pattern length scale\label{subsec:evidence_pattern_length_scale}}
\begin{figure*}
\centering
\includegraphics[width=\textwidth]{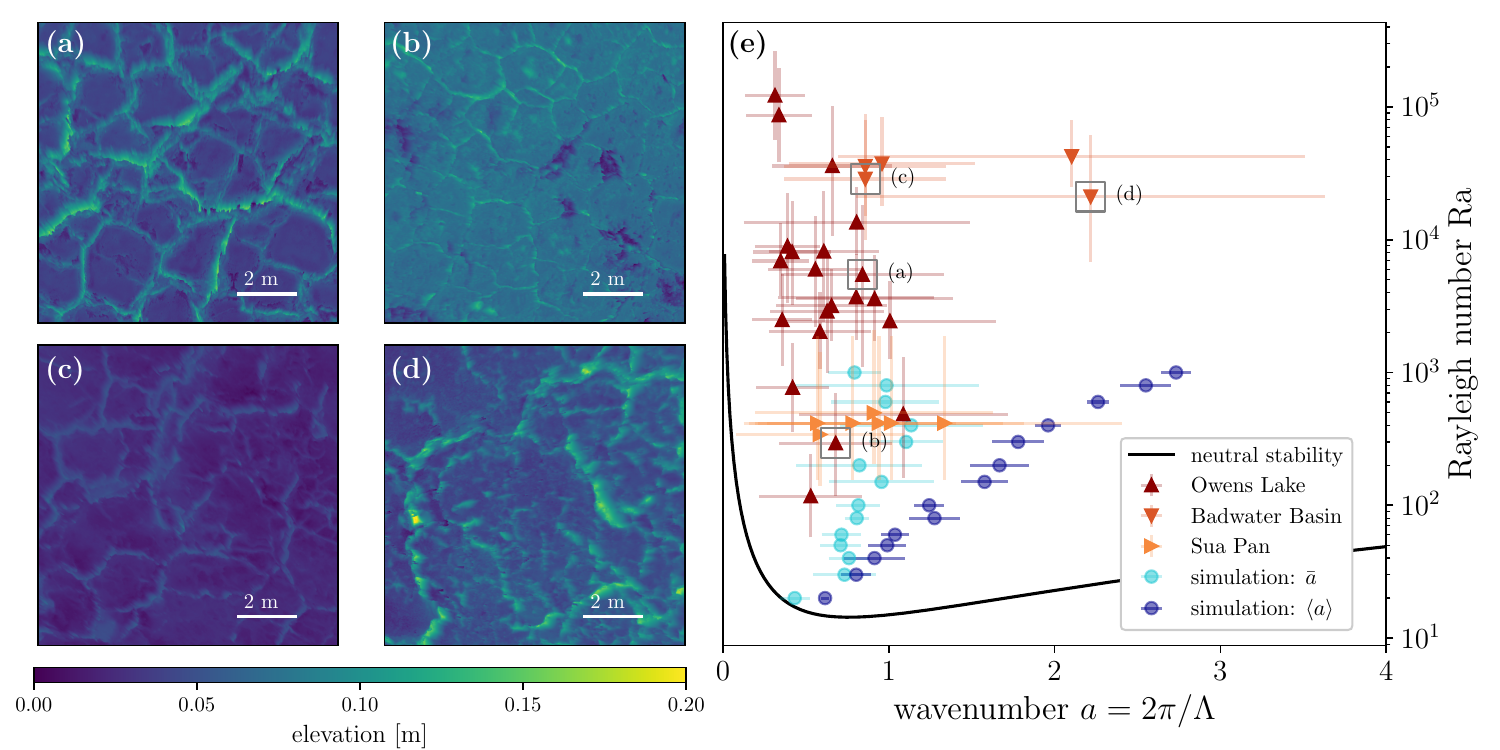}
\caption{Field observations of the length-scale selection of salt polygons.  Panels \label{fig:fig8} \textbf{(a)} to \textbf{(d)} show the surface relief recorded at several field sites. Panel \textbf{(e)} maps data from the field onto the stability diagram of porous media convection in a salt pan. The neutral stability curve (black line) is the theoretical boundary above which an evaporating stratified pore fluid is unstable to perturbations of wavenumber $a$.  Various triangles show measurements at Owens Lake, Badwater Basin and Sua Pan, with letters indicating data from the sites in \textbf{(a--d)}. Turquoise and blue dots show the average and dominant wavenumber measured in simulations, respectively (data from Fig.~\ref{fig:fig4}).}
\end{figure*}

If groundwater convection leads to preferential locations for salt precipitation, and from thence to salt crust patterning, then the convective cells and crust polygons should have similar length-scales and patterns.  We presented model predictions for the length-scale selection of convection in Section~\ref{sec:buoyancy_driven_convection_as_a_mechanism_for_pattern_formation} B.  To test these predictions in the field, we measured the surface relief of the crusts at multiple sites from all three dry lakes using a terrestrial laser scanner (see Appendix~\ref{app:field}).  The crusts display the typical polygonal patterns of salt lakes, as shown in Fig. \ref{fig:fig8}(a--d), with further data in \cite{lasser2020surface,NIELD2015}.  From each scan we extracted a characteristic wavelength for the pattern, based on the dominant frequency of the power spectrum, using the methods reported in~\cite{NIELD2015}.  Full results are given in the Appendix, Table~\ref{tab:tab4}.  In all cases the pattern wavelengths ranged from $\lambda = 0.53\pm 0.20\,$m to $3.02\pm 1.40\,$m.  These were converted into dimensionless wavenumbers $a = 2\pi L / \lambda$, using the characteristic lengthscale $L = \varphi D/E$ as calculated for each lake ($E$ from Section~\ref{sec:evidence} A, $\varphi$ and $D$ from Section~\ref{sec:evidence} C).   The results are summarized in Fig.~\ref{fig:fig8}(e), which shows the Rayleigh number and wavenumber for each site studied.

We found that the wavenumbers characterizing the surface crust patterns in the field are clustered near the critical wavenumber for convection, \mbox{$a_\mathrm{c}=0.76$}, and largely independent of Ra.  {In Section~\ref{subsec:scale_selection}, we predicted how the statistically steady state of subsurface convection should look. In this state the average wavenumber increased slowly with Ra, by about a factor of four from Ra = 20 to 1000.  This was attributed to the proliferation of small, transient features, superimposed on more stable convection cells.  The dominant wavenumber, given by the peak of the power spectrum, showed little-to-no dependence on Ra, and was argued to better represent the long-term behavior of the convection cells.  As shown in Fig.~\ref{fig:fig8}(e), although there is large scatter in the data, the wavelengths of the crust pattern broadly match the wavelengths of the dominant features of the sub-surface convection expected in a statistically steady state.} 


Furthermore, for the field sites there should be ample time for the convective dynamics to develop towards the steady-state condition.  In the model, the timescale most appropriate to characterize the dynamics of convection is $T/\epsilon$, where $T = \varphi^2 D/E^2$ describes how the boundary layer would develop in the absence of convection, and $\epsilon = (\mathrm{Ra}-\mathrm{Ra_c})/\mathrm{Ra_c}$ accounts for the increased vigor of convective motion as the driving forces increase (see also~\cite{LASSER2021}).  For the relevant parameter values of Owens Lake, $T/\epsilon$ is just over one day.  Alternatively put, one year of evolution at Owens Lake, for a site with a representative value of $\mathrm{Ra} = 3700$, would represent a rescaled time of $\widehat\tau = 350$.  This is long compared to the timescale of $\widehat\tau = 30$ that we used to characterize the steady-state conditions in Figs.~\ref{fig:fig4} and \ref{fig:fig8}.

{Finally, as shown in Fig.~\ref{fig:fig8}(a--d), polygon ridges are relatively narrow features. To characterize their size, we thresholded our surface relief maps using Otsu's method~\cite{Otsu1979}. The relative area occupied by ridges, measured in this way, was 24$\,\pm\,8\%$ (mean $\pm$ standard deviation); this may be a slight overestimate, due to occlusion of the TLS behind larger ridges. In the 3D simulations the comparable features are downwellings, which we characterized as areas of positive surface salinity flux at $\widehat\tau = 30$. These covered 11--18\% of the surface; narrower features were seen at higher Ra, but the range of Ra explored is small. Although the salt ridges can be relatively complex, involving breaking and thrusting for example, the convective model qualitatively captures the essential asymmetry between the narrow ridge features and the flat pans between them.}

In summary, the pattern wavelength we measured for salt polygons at a range of sites across three different salt lakes is thus found to be consistent with the length scale expected for buoyancy driven convection going on in a statistically steady state. Simulations of the dynamics in a three-dimensional domain (see Fig.~\ref{fig:fig3}(d)) also show that the convective mechanism can produce patterns of closed polygonal shapes at this lengthscale.  These patterns are qualitatively similar to the salt polygons observed in the field (Fig.~\ref{fig:fig8}(a--d)) {with relatively narrow features delimiting the larger polygonal shapes}.

\subsection{Environmental variation \& remaining challenges}\label{subsec:evidence_robustness}
To this point, we have focused on the effects of varying  environmental conditions as they contribute to the Rayleigh number and natural lengthscale of the convection problem. We showed that a wide range of measured field conditions should lead to convection, and to convection cells with a size and pattern that is consistent with the observed surface crust patterns; {faster crust growth, and thicker accumulations in ridges, was shown to correlate with the downwellings of the subsurface convection.} In this section, we will briefly look at and discuss other factors, including the salt chemistry, crust thickness and features like cracks and crust wrinkling.  {We will also discuss some of the remaining assumptions and challenges of the convective model, in explaining crust patterning.}

In addition to the soil, we characterized the composition of the salt crust at Owens Lake and Badwater Basin (see Appendix~\ref{app:chemistry} for methods; data deposited at~\cite{CHEMICAL}). Both sites are predominantly halite, although Owens Lake additionally has significant concentrations of carbonates and sulfates~\cite{CHEMICAL, TYLER1997}.  Between all the field sites we also observed salt crusts with thicknesses from below a centimeter to approximately $30\,\mathrm{cm}$. Across these differences in soil and salt crust composition there is a similar appearance of the salt polygons.  This agrees with a mechanism like subsurface convection, which is relatively insensitive to salt chemistry.  It would, conversely, present difficulties to any mechanism that was reliant on specific salts, such as mirabilite, or which varied strongly with crust thickness.  

\begin{figure}
\centering
\includegraphics[width=0.47\textwidth]{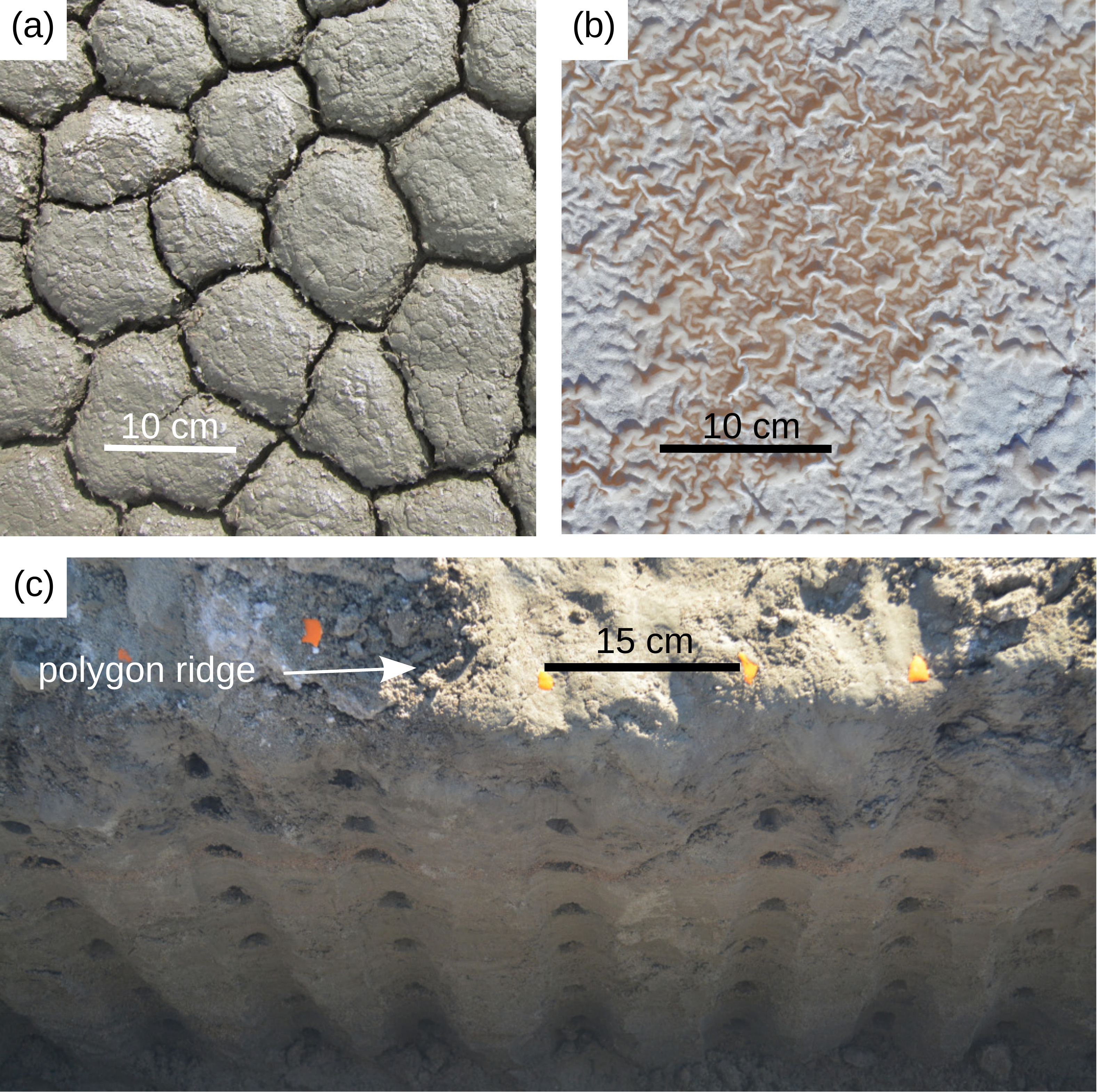}
\caption{\label{fig:fig9} Additional surface and subsurface features at Owens Lake. \textbf{(a)} Small-scale hexagonal mud cracks could be seen in the soil about $10\,$m away from one of our field sites. \textbf{(b)} Wrinkles were also observed in the salt crust adjacent to one of the polygons that served as a field site. \textbf{(c)} At a trench site the soil layering was exposed.  The surface position of a polygon ridge is indicated by the white arrow, and holes mark the sampling locations. The horizons in the soil beneath the polygon are visible and undisturbed (see for example the reddish layer near the third sampling hole from the top).}
\end{figure}

Other surface features, including cracks and wrinkling or buckling, are also associated with salt crusts \cite{CHRISTIANSEN1963,KRINSLEY1970}. While studying Owens Lake, we observed well-ordered hexagonal desiccation cracks in crust-free mud with approximately 10 cm spacing, as shown in Fig.~\ref{fig:fig9}(a).  These appeared at the crust-free edge of a pan that was otherwise covered in salt ridges (with a spacing of 1--2$\,\mathrm{m}$).  The hexagonal crack patterns can be explained as the result of the intermittent and episodic cracking of the soil, as water levels fluctuate \cite{GOEHRING2013}.  However, they form at a very different scale to the salt ridges, and the cracks develop as localized depressions in the ground, rather than upthrust ridges. Further, despite their presence near the salt crust, there were no signs of preferential precipitation of salt in the cracks. We also observed surface buckling of the salt crust at Owens Lake, as shown in Fig.~\ref{fig:fig9}(b) (see also supplementaal video S1 on dates 05/07 through 05/11).  These features appeared when the crust was solidifying, and may result from the stresses induced by phase changes or the addition of salt mass into the crust.  However, the scale of their features is, again, much smaller than those of the dominant polygonal ridges in the crust, and the patterns formed are also different.  Finally, we note that throughout our core samples and trench sites, as in Fig.~\ref{fig:fig9}(c), the soil horizons remained normal and undisturbed by the presence of surface ridges.  This would not be the case if deep fractures were occurring along with the ridge formation (as in \textit{e.g.} ice-wedge formation in permafrost terrain~\cite{LACHENBRUCH1962}).

{More generally, we have treated the salt crust as two-dimensional, of nominal thickness.  This is a necessary simplification, given the novelty of the hypothesis advanced here. However, there will be some internal dynamics within the salt crust itself, beyond the scope of our model.  For example, dissolution will be favored at the base of the crust, in contact with groundwater, and precipitation at the top, where evaporation occurs.  Similar observations have been made in playa~\cite{NACHSHON2018} and in lab~\cite{DAI2016,LISCANDRU2019} studies of efflorescence with cultural heritage applications.  Dissolution-precipitation dynamics raises porosity in crusts, leading to a `fluffy' texture, trapped air pockets, self-lifting effects~\cite{NACHSHON2018,DAI2016,LISCANDRU2019}, and a lower threshold for dust entrainment~\cite{Reynolds2007}.  We accounted for the high crust porosity while calculating growth rates in section~\ref{subsec:evidence_surface_salt_flux}, but a more nuanced approach may give further insight, including into dust models.}  

{There is also likely to be an element of positive feedback between the growth of the salt crust, and the dynamics of the convective plumes.  Our simulations showed that, without such a feedback, the positions of downwelling features slowly fluctuate in space, although this tendency is weaker in 3D than in the 2D simulations.  A feedback cycle, where evaporation rates are affected locally by the presence of the ridges, can act to stabilize the locations of the plumes.  This has been demonstrated~\cite{LASSER2021} by simulations in which downwellings tend to move towards, and stay at, locations with lower evaporation rates.   We have also measured temperature and relative humidity records from Owens Lake~\cite{LASSER2021}, which show that evaporation rates should be slower at the ridges, compared to the centers of polygons, in a way consistent with such a feedback mechanism.  Further model development along these lines would be promising.}

{Incorporating feedback into crust evaporation may also help refine predictions of wavelength selection.  The scale of salt polygons matches the dominant wavelength of our simulations, which we argued to be representative of a long-time integration of the more stable convection pattern.  The effect of the smaller-scale, transient `microplumes' remains uncertain.  As captured by the average wavenumber, these are more prevalent at higher Ra, but their faster dynamics (changing visibly over simulation times of $\sim 0.1$, or about two weeks in real terms) mean that they should only weakly contribute to the surface pattern, which develops over months.  This prediction could be checked in simulations that modulate surface evaporation, based on tracking of the total amount of salt accumulating at the surface over time.}

{As a final point, one may also consider thermal contributions to buoyancy, and phenomena like double-diffusive convection.  A typical 10$^\circ$C diurnal variation~\cite{LASSER2021} will change water density by about $\Delta\rho_T$ = 1 kg m$^{-3}$, compared to the $\Delta\rho_S\simeq$ 200 kg m$^{-3}$ density contrast due to solutes at Owens Lake. The buoyancy ratio, $N=\Delta\rho_S / \Delta\rho_T\simeq 200$, means that the driving forces and speeds of thermally-driven flows will be reduced by a factor of 1/$N$, compared to solute-driven flows (see \textit{e.g.}~\cite{Mojtabi2005} for review of double-diffusive flow in porous media).  Such effects can, therefore, be assumed to be negligible.}   


\section{Discussion and Conclusion}
Salt deserts, playa and pans are a common landform important for climate balances such as dust, energy and water, and express a rich repertoire of patterns and dynamics. Here we have shown that, in order to model and understand the surface expression of such deserts, insight is gained by considering the crust together with the subsurface dynamics. In particular, we have shown how the emergence of regular salt polygons, which are a common salt crust pattern, can result from their coupling to a convection process in the soil beneath them. The existence of salinity-driven convection in salt pans is itself already an anticipated finding (see \textit{e.g.}~\cite{WOODING1997b}).  Furthermore, a salinity gradient below the surface, strong enough to drive such convection, has been measured at Owen's Lake~\cite{TYLER1997} and strong horizontal salinity gradients indicative of convective motion have been shown to exist in sabkhas~\cite{VANDAM2009,STEVENS2009}.
What we have demonstrated is that this convection process {develops into a statistically steady state characterized by large-scale convection cells with a} wavelength of a few meters, after a few weeks of evolution under typical field conditions. We further showed how this can explain the remarkably consistent appearance, growth rates, polygonal shape, narrow ridges, and representative wavelengths of the salt crust patterns observed in different parts of the world, including our field sites at Sua Pan, Badwater Basin and Owens Lake.

{Our findings help to elucidate the relationships between subsurface drivers and the development of surface crust patterns. Better knowledge of crust development can improve dust emission models~\cite{Klose2019,Tollerud2014,Reynolds2007}, including quantifying the connection of salt pan hydrology to controls on dust emission~\cite{Sweeney2016} such as those used at Owens Lake~\cite{GROENEVELD2010,GROENEVELD2013}.  As further context, dusts and aerosols remain one of the larger uncertainties in modelling climate sensitivity~\cite{IPCC6}, with \textit{e.g.} feedback from soil biocrust degredation recently predicted to contribute half as much as direct anthropogenic aerosol emissions by 2070~\cite{Caballero2022}.  While the surface variability of crusted surfaces can be probed using remote sensing~\cite{Milewski2020}, our model is a first step to improving predictions of how subsurface hydrologic variability links to surface crust variability.  This knowledge is increasingly important as saline lakes like the Dead Sea or Great Salt Lake are shrinking~\cite{Wurtsbaugh2017} and their newly exposed beaches are candidate areas for salt crust formation and dust emission.}

To establish the connection between surface features and subsurface flows, we demonstrated consistent results from theoretical and numerical modeling, analogue experiments and field studies. In contrast to previous explanations of crust patterns~\cite{CHRISTIANSEN1963, KRINSLEY1970}, this model is able to explain the robustness of the pattern length scale by considering the statistically {steady state of porous media convection,} based only on measured environmental parameters. In fully three-dimensional simulations, the convective dynamics were also shown to give rise to closed-form polygonal shapes.  {More importantly, it is able to predict a suite of quantitative details, such as the timescales required for convection to set in, the rates of crust and ridge growth, the positions of the ridges above salt-rich plumes, and the relative narrowness of the ridge features.} At the downwellings the salinity is higher and therefore the salinity gradient between the crust and the underlying fluid is weaker (compare sketch of Fig.~\ref{fig:fig2} to measurements in Fig.~\ref{fig:fig7}). As salt transport is a balance of advective and diffusive processes, this will lead to an increased rate of salt precipitation above downwelling features, contributing to the growth of ridges at the boundaries of convection cells. After the initial emergence of ridges, the growth process might be bolstered by feedback mechanisms such as a modulation of the evaporation rate by the presence of ridges, cracks or surface wicking phenomena.

As such, our results show how salt polygons are part of a growing list of geophysical phenomena, such as fairy circles~\cite{JUERGENS2013}, ice wedges~\cite{SLETTEN2003}, polygonal terrain \cite{Kessler2003} and columnar joints~\cite{GOEHRING2013}, which can be successfully explained as the result of the instability of a dynamical process. 

\section{Data and code availability}
Field site labels and locations as well as site-by-site data for Fig.~\ref{fig:fig8}\textbf{(e)} are given in the appendix Table~\ref{tab:tab4}.  Full data sets for the soil particle sizes~\cite{GRAINSIZES}, surface height profiles~\cite{TLS}, salt chemistry~\cite{CHEMICAL}, subsurface salinity distribution and pore water density~\cite{CONCENTRATION} as well as images describing the US field sites~\cite{IMAGES} are available under a CC-BY-SA 4.0 license at PANGAEA.

Code for the 2D simulation is available under an MIT license on GitHub, \url{doi:10.5281/zenodo.3969492}.

All scripts, data sets and images used to produce the figures in this work are available at \url{https://doi.org/10.17605/OSF.IO/KJDWF}.

\section{Author contributions}
JL, ME and LG wrote the original draft of the manuscript; all authors reviewed the manuscript.

JL participated in fieldwork at Owens Lake and Badwater basin, analyzed the soil salinity and grain size distributions, performed numerical experiments of the 2D simulation, performed the Hele-Shaw experiments, analyzed all data except for the 3D simulations and TLS measurements and created the data visualizations. 

JMN participated in fieldwork at Owens Lake, Badwater Basin and Sua Pan and performed and analyzed the TLS measurements.

ME developed the theory of convective dynamics and programmed the 2D simulation.

VK performed the X-ray analysis, supervised JL for the grain size measurements and provided feedback for the soil composition and salt species analysis.

GFSW participated in fieldwork at Sua Pan.

MRT performed numerical experiments and analyzed the data from the 3D simulations.

CB supervised MRT, programmed the 3D simulation and provided feedback for the theoretical model.

LG conceptualized the study, developed the theory of convective dynamics, supervised JL and ME, performed the water density measurements and participated in fieldwork at Owens Lake and Badwater Basin.

\begin{acknowledgments}
We thank Grace Holder (Great Basin Unified Air Pollution Control District) for support at Owens Lake, the U.S. National Park Service for access to Death Valley (Permit DEVA-2016-SCI-0034) and Antoine Fourri\`ere and Birte Thiede for their work on preliminary experiments on convection. TLS processing used the Iridis Southampton Computing Facility. MRT was supported by the Leeds-York-Hull Natural Environment Research Council (NERC) Doctoral Training Partnership (DTP) Panorama under grant NE/S007458/1. Sua Pan work was funded by Natural Environment Research Council (NE/H021841/1), World University Network and Southampton SIRDF (Strategic Interdisciplinary Research Development Funds) and enabled by Botswana Ministry of Environment, Wildlife, and Tourism (Permit EWT 8/36/4 XIV) and Botswana Ash (Pty) Ltd. 
\end{acknowledgments}

\appendix

\section{Numerical simulation}\label{app:simulation}

The three-dimensional (3D) simulations use a pseudo-spectral element method where the horizontal directions, over which the numerical domain is periodic, are treated using fast Fourier transforms, and the vertical (bounded) direction is decomposed into elements, each of which is treated using Gauss--Lobatto--Legendre quadratures (see \textit{e.g.}~\cite{CanutoBook}). The time-derivative is approximated using a second-order backward differentiation formula and the nonlinear term extrapolated using a Taylor expansion of second order. The coefficients of the resulting schemes can be found in \cite{Karniadakis91}.  All simulations were carried out on a domain of depth 10 and horizontal area of 24$\pi \times 24\pi$.  

{To compute wavenumbers using two-dimensional simulations (\textit{cf.} Fig.~\ref{fig:fig4}), we used the same numerical solver and imposed the solution invariance in one of the horizontal directions.  For the two-dimensional (2D) simulations of wavenumber selection (Fig.~4), we used the same numerical solver and imposed the solution invariance in one of the horizontal directions. These simulations have horizontal domain size $12\pi$ and depth 10 or 20. Pattern wavenumbers were calculated from the salinity at the gridpoint closest to $Z = 1$ as in 3D, but without radially-averaging the power spectra.  The dominant wavenumber was identified as the mode with the highest power spectral density.  All other two-dimensional (2D) simulations were carried out using a stream-function-vorticity approach based on Refs.~\cite{RIAZ2003,CHEN1998} with a detailed implementation described in Ref.~\cite{LASSER2021}.} 

\section{Field data collection}\label{app:field}
\begin{figure}
\centering
\includegraphics[width=0.47\textwidth]{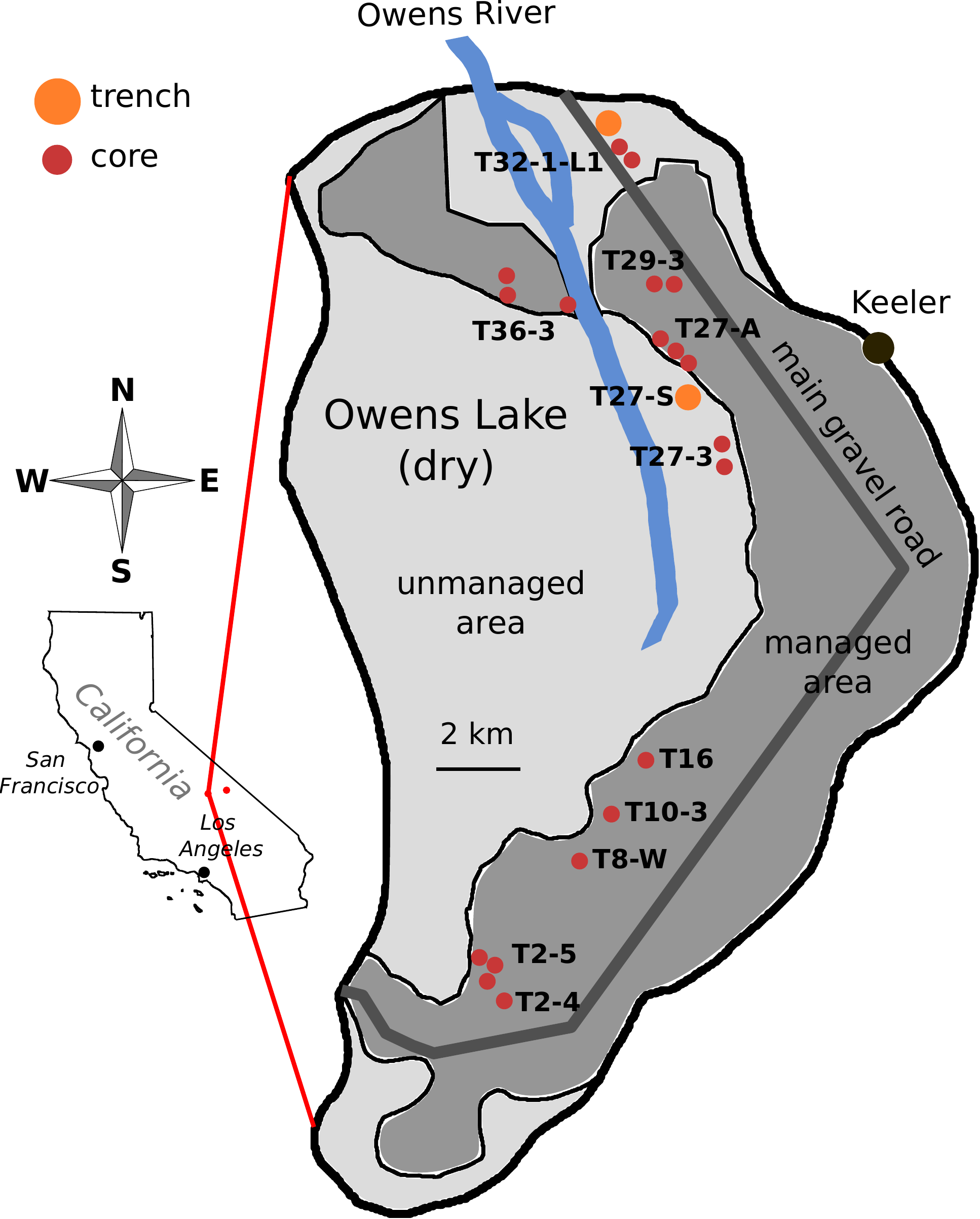}
\caption{\label{fig:fig10}Schematic map of Owens Lake (CA, USA) with locations of soil core (red) and trench (orange) sites marked. Labels give the names of local dust management areas.}
\end{figure}

Field work was conducted at Owens Lake and Badwater Basin in November 2016 and January 2018; see \textit{e.g.}~\cite{HOLLET1991,HUNT1966} for geological descriptions of the area.  The Owens Lake sampling sites are indicated in Fig.~\ref{fig:fig10}. At Badwater Basin five sites were visited $\sim$500$\,$m south of the main tourist entrance to the playa.  For both locations field methodology and data collection are fully documented in a separate data description publication~\cite{lasser2020surface}.  

At Owens Lake we use field site labels referring to surface management cells of the dust control project there~\cite{LADWP2010}. These labels link either to managed cells or to unmanaged areas in the direct vicinity of a managed cell. Labels typically start with ``TX-Y", where X is a number and Y is a number or letter. The first number refers to the water outlet taps along the main water pipeline that crosses the lakebed south to north and is used to irrigate management cells.  Numbers/letters after the hyphen refer to subregions branching from the same tap.  For some sites we investigated more than one polygon. This is indicated with brackets, \textit{e.g.} T27-A (3) is the third polygon investigated at site T27-A, which corresponds to the Addition (A) region of the management cell next to the 27$^\text{th}$ tap.  Bracketed numbers are used for the sites visited at Badwater Basin.

To evaluate the profile of salt concentration below polygons, at most sites soil cores (4$\,$cm Dutch gouge auger) were taken to a depth of up to $1\,$m. The soil showed normal bedding (see also Fig.~\ref{fig:fig9}(c)), indicative of sedimentation following flooding. Samples were collected from each visible soil horizon, or with a vertical resolution of $\Delta z = $10--15$\,$cm. 

Trenches were also dug at sites T27-S (Site 1 in Figures~\ref{fig:fig7} and~\ref{fig:fig11}) and T32-1-L1 (3) (Site 2), in order to take samples along cross-sections below salt polygons.  
The trenches were dug about 200$\,$cm in length, 40$\,$cm in width and down to a water table of $\sim$70$\,$cm.  Soil samples were taken from a freshly cleaned trench wall in a grid pattern with spacings of $\Delta x = 15\,$cm and $\Delta z = 10\,$cm. An example of such a trench along with sampling locations is shown in Fig.~\ref{fig:fig9}(c). The samples had an average volume of approximately $10\,$ml and were taken using a metal spatula, which was cleaned with distilled water and dried before each use. The samples were a mixture of soil with a grain size of medium sand to clay, water and salt (both dissolved and precipitated). After collection, samples were immediately stored in air-tight containers, which were sealed with parafilm. 

To evaluate the density difference, $\Delta \rho$, we collected pore water samples at Owens Lake from eight sites, including liquid taken from directly below the surface (in cases where the water table was at or within 10 cm of the surface), and at a depth of about $1\,$m.  

Samples from Sua Pan were collected during a different field campaign, reported in Ref.~\cite{NIELD2015}; site labelling follows that publication.  Sediment samples at Sua Pan were collected 2$\,$cm below the crust in August 2012. These were double bagged and subjected to grain size analysis only.

At all three dry lakes, surface height maps were collected with a Leica Terrestrial Laser Scanner (TLS); a P20 model was used at Owens Lake and Badwater, and a Scanstation at Sua Pan.  The scanner head was positioned at least $2\;$m above the playa surface and scans were performed before the surface was disturbed by sampling. 

GPS locations of all sites are provided in Tab.~\ref{tab:tab4}.

\section{Soil characterization}\label{app:grain_size}
Soil samples from Owens Lake, Badwater Basin and Sua Pan were analyzed to determine their distribution of grain sizes. This analysis was performed using the soil remaining after a sample's salinity was determined (Appendix~\ref{app:salinity}). Soil grain size distributions were measured by laser particle sizer (Coulter LS 13 320), from which the Sauter diameter (the mean diameter, respecting the soil's specific surface area \cite{SAUTER1928}) was calculated.  For each site a representative $d_S$ is calculated as the average Sauter diameter of all soil samples (for trenches, one sample per depth) from that site. For Sua Pan, only samples from sites B7 and L5 were available; $d_S$ for the other five sites is estimated as the mean of the measured values at these two sites.   All grain size data is deposited at Ref.~\cite{GRAINSIZES}.

Soil porosity has previously been measured to be around $\varphi\approx 0.70\pm0.02$~\cite{TYLER1997} at Owens Lake. Because of lack of similar measurements at Badwater Basin and Sua Pan, we used the value measured at Owens for calculations of $\kappa$ and Ra at these sites.  For each site a permeability is then calculated based on the Sauter diameter and porosity, as $\kappa=0.11\varphi^{5.6}d_S^2$~\cite{GARCIA2009}.

The Sauter diameter $d_S$, permeability $\kappa$ and Rayleigh number Ra  for all sites investigated are provided in Table~\ref{tab:tab4}.  Uncertainty ranges for Ra were calculated as systematic errors based on standard errors of all the input environmental parameters. 

\section{Evaporation rate}\label{app:evap}
{Groundwater evaporation rates are taken from the literature. At Owens Lake, Tyler \textit{et al.}~\cite{TYLER1997} combined lysimeter, Bowen ratio and eddy correlation methods, with repeat seasonal measurements conducted well after any rainfall.  The value of $E = 0.4 \pm 0.1 $ mm/day captures the range of measurements at their NFIP site, near our northern sites at the lake. At Badwater Basin, DeMeo \textit{et al.}~\cite{DEMEO2003} measured evaporation by Bowen ratio and eddy correlation methods, with continuous monitoring over several years.  The value of $E = 0.3 \pm 0.1 $ mm/day is taken from their two playa sites south of Furnace Creek, and the 2001 conditions most consistent with the wetter season preceding our field work. At Sua Pan we use $E=0.7\pm0.5\,$mm/day, estimated by remote sensing and energy balances \cite{BRUNNER2004}, and wind tunnel experiments~\cite{NIELD2016}.}

\section{TLS data processing}\label{app:tls}
TLS Scan data was processed following Ref.~\cite{NIELD2013}. Data were first gridded into a digital elevation map (DEM) with a lateral resolution of $1\,$cm and a vertical resolution of $0.3\;$mm. Dominant frequencies of surface roughness, which we use here for the pattern wavenumber $a$ (and wavelength $\lambda = 2\pi/a$ as reported in Table~\ref{tab:tab4}), were then quantified using the $90^\text{th}$ percentiles determined with the zero-upcrossing method from the DEMs~\cite{NIELD2015}. TLS data from the US field sites are deposited at Ref.~\cite{TLS}.

\section{Salinity and density measurements}\label{app:salinity}
Owens Lake and Badwater Basin soil samples were analyzed to determine the amount of salt in their pore fluid. These samples had been sealed immediately after collection, to preserve their water content.  After unsealing, each sample was first transferred to a crystallizing dish and weighed, to give a combined mass of sand, salt and water. It was then dried at 80$^\circ$C until all moisture had visibly vanished, or for at least 24~h, and re-weighed to determine the mass of the (evaporated) water. Next, it was washed with 50$\,$ml of deionized water to dissolve any salt, allowed to sediment for 24 hours, and the supernatant liquid was collected in another crystallizing dish. After two such washings the remaining soil and the recovered salt solution were dried and weighed.  Measurement uncertainty is based on the difference between the initial sample mass and the sum of the separated water ($m_w$), salt ($m_s$) and soil masses.  This gave a direct measure of the mass fraction of salt in solution as $C = m_s/(m_w+m_s)$, reported as a weight \% (wt.\%).

Owens Lake pore water samples were analyzed to determine their density using a vibrating-tube densitometer (Anton Paar DMA4500). The near-surface (0--10 cm depth) groundwater density was consistently $1255\pm8\,$kg\,m$^{-3}$, while water from 70--100 cm depth had a density of $1050\pm2\,$kg\,m$^{-3}$. These values are broadly consistent with chloride concentration profiles previously measured at Owens Lake \cite{TYLER1997}. We note that thermal effects on the groundwater density will be comparatively negligible, as the mean annual variation in temperature at Owens Lake will allow for a density change of, at most, 5\,kg\,m$^{-3}$.  Similarly, the solubility of halite in water would change by less than 5\,kg\,m$^{-3}$, seasonally.  
Data from salt concentration and pore water density measurements are deposited at Ref.~\cite{CONCENTRATION}.

\section{Salt chemistry and diffusivity}\label{app:chemistry}
Pore water samples from selected sites at Owens Lake and Badwater Basin were analyzed to determine the dominant salt species present. This analysis was performed on the dried salts remaining after the salinity measurements (Appendix~\ref{app:salinity}). Mineral identification was performed by quantitative X-ray powder diffraction analysis (Philips X'Pert MPD PW 3040). Samples from Owens Lake showed a mixture of salts with $53\pm7\,$wt.\% sodium chloride and $30\pm5\,$wt.\% hydrated sodium sulphate (mirabilite). Other minerals, such as natrite, sylvite and burkeite, were variously present at less than 10 wt.\%, each.  As such, in solution Na$^+$ and Cl$^-$ ions predominate ($>$70\% by mass), with SO$_4^{2-}$, CO$_3^{2-}$ and K$^+$ ions present in descending order of significance.  All data are deposited at Ref.~\cite{CHEMICAL}.

Based on the ionic species found in the pore water, we estimate an average aqueous diffusivity of \mbox{$D^*=1.37\times10^{-9}\,\text{m}^2\,\text{s}^{-1}$} from measurements of ternary mixtures of the two primary salts~\cite{ANNUNZIATA2000}, using a weighted average of the mole ratios of their main-term diffusion coefficients. Accounting for the tortuosity, $\theta$, of the porous medium, we then calculate an effective diffusion coefficient $D=D^*/\theta^2 = (1.00\pm 0.24)\times 10^{-9}\,$m$^2$\,s$^{-1}$ following \cite{BOUDREAU2011}, where we estimate $\theta^2=1-\ln(\varphi)$, as in \cite{BOURDEAU1996}.

\section{Fitting the salt-rich boundary layer}\label{app:salt_flux}
\begin{figure}
    \centering
    \includegraphics[width=0.47\textwidth]{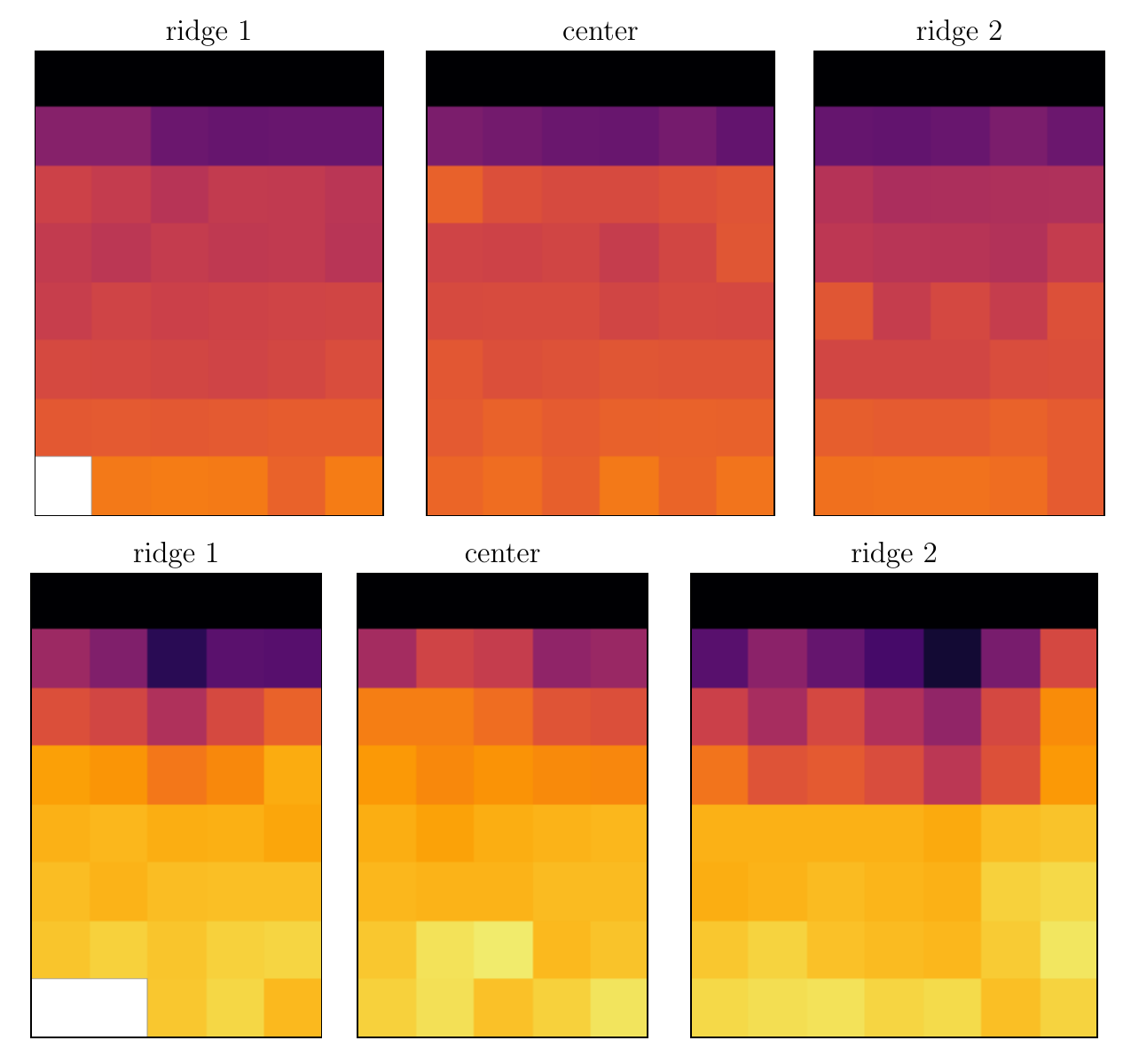}
    \caption{Division of data from the two trench sites into two ridge areas and one center area each.  The horizontal sample spacing is $15\,$cm. White areas indicate missing samples.}
    \label{fig:fig11}
\end{figure}

\begin{figure*}
    \centering
    \includegraphics[width=\textwidth]{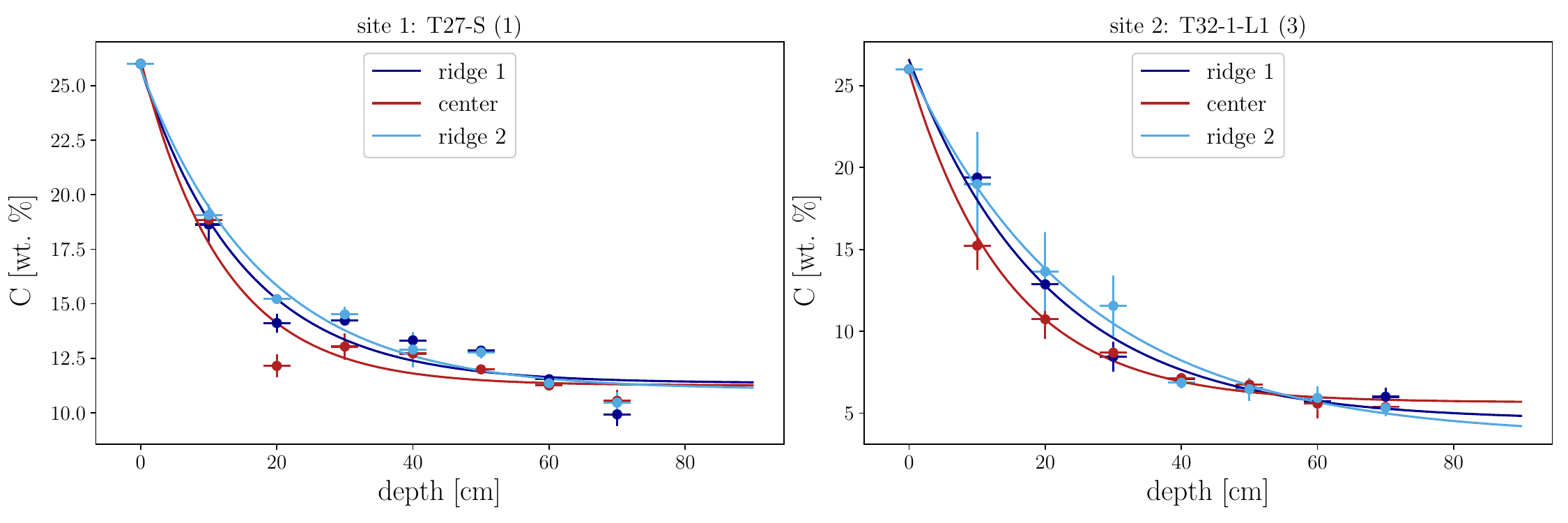}
    \caption{Horizontally-averaged salt concentration, $C(z)$, at the ridge (blue) and center regions (red) of the two trench sites (dots) along with exponential fits (lines). Error bars indicate standard deviations of the concentrations, along with a sample position uncertainty of $\pm 2\,$cm.}
    \label{fig:fig12}
\end{figure*}

We used the distributions of salt at the two trench sites to estimate the near-surface gradients in salt concentration, and from this the surface salt flux, $J_{c, z}$.  For this, we divide the grid of measurements at each site into ridge and center areas--as shown in Fig.~\ref{fig:fig11}. Ridge areas cover about $40\,$cm on either side of a polygon ridge. The centers include samples from the areas between the ridges, and were approximately $70\,$cm wide.  An exponential of the form \mbox{$C(z) = C_\mathrm{bkg} + (C_\mathrm{sat} - C_\mathrm{bkg})e^{z/L'}$} was then fit to the horizontally-averaged salt concentrations ($C$, wt.\%) of each respective area, using an error-weighted fitting of the data.  Here, $L'$ is the effective thickness of the salt-rich boundary layer, $C_\mathrm{bkg}$ is a background salt content of the pore fluid at depth and $C_\mathrm{sat}$ is the salt content of fluid at the surface.  The fits are demonstrated in Fig.~\ref{fig:fig12} and output values for the fit parameters are given in Table~\ref{tab:tab2}, with associated errors.  As a consistency check, we note that all fitted $C_{\mathrm{sat}}$ values agree with the mass fraction of a saturated NaCl solution (26.3 wt.\% at 10$^\circ$C \cite{Solubility47}), which is by far the most predominant dissolved salt in the pore water at Owens Lake (see Appendix~\ref{app:chemistry}). Following fitting the salt concentrations, in terms of a weight per unit volume, were calculated as $c = \rho C$, assuming densities $\rho$ based on NaCl solutions at 10$^\circ$C \cite{PERRY2008}.

\begin{table}
    \centering
    \begin{tabular}{ccccc}
         site & area & $L'$ [cm] & $C_\mathrm{bkg}$ [wt.\%] & $C_\mathrm{sat}$ [wt.\%]\\
         \hline
         site 1 & ridge 1 & $15.1\pm 3.3$ & $11.4\pm 0.8$ & $25.8 \pm 2.1$\\
         site 1 & ridge 2 & $17.8\pm 2.6$ & $11.1\pm 0.6$ & $25.8 \pm 1.5$\\
         site 1 & center & $12.1\pm 2.6$ & $11.3 \pm 0.7$ & $26.2 \pm 2.0$\\
         \hline
         site 2 & ridge 1 & $20.3\pm 2.6$ & $4.6\pm 0.9$ & $26.6 \pm 2.0$\\
         site 2 & ridge 2 & $25.3\pm 3.4$ & $3.6\pm 1.1$ & $26.1 \pm 2.2$\\
         site 2 & center & $14.5\pm 1.0$ & $5.7\pm 0.4$ & $25.8 \pm 0.9$\\
         \hline
    \end{tabular}
    \caption{Boundary layer thickness $L'$, along with the salt content (wt.\%) of background ($C_\mathrm{bkg}$) and saturated ($C_\mathrm{sat}$) pore fluid, from exponential fits to the salt concentration distributions below the center and ridge areas of two trench sites.}
    \label{tab:tab2}
\end{table}

\section{Experimental methods}\label{app:experiments}

Experiments were performed in 40\,$\times$\,20\,$\times$\,0.8~cm (width$\times$height$\times$gap spacing) Hele-Shaw cells.  As summarized in Table~\ref{tab:tab3}, these were filled with glass beads (Sigmund Lindner GmbH) with diameters of 0--20$\,\mu$m to 200--300$\,\mu$m, $d_S = 1.9\pm 0.2\,\mu$m to $264\pm 25\,\mu$m, and $\varphi = 0.37\times 10^{-2}$ to $\varphi = 0.21\times 10^{-2}$. The permeability of the bead packs was evaluated by flow-through experiments.  The base of each cell was connected to a reservoir containing a $50\,$kg\,m$^{-3}$ solution of NaCl, such that $\Delta\rho =162\,$kg\,m$^{-3}$ (compared to a saturated salt solution \cite{Solubility47}). This reservoir maintained a fluid-saturated pore space between the beads. Evaporation at the top of the cells was controlled and enhanced by overhead heating and air circulation, and could be controlled from $E \approx 1\,$mm/day to $10\,$cm/day.  Assuming a kinematic viscosity of $\mu = 10^{-3}\,$Pa s, these conditions allowed for experiments from Ra $= (7\pm 2) \times 10^{-3}$ to Ra $= (1.3\pm 0.3) \times 10^3$. The experiments with $\mathrm{Ra} < \mathrm{Ra}_\mathrm{c}$ did not show any convective dynamics over a period of three months, while all experiments with $\mathrm{Ra} > \mathrm{Ra}_\mathrm{c}$ showed convection. 

Visualization of the convective dynamics in the cells (see Supplemental movie S4)  was accomplished by  intermittently injecting $2$--$4\,$ml of dyed saline solution through a thin tube embedded in the cell. The dye was then advected by the flows inside the cell over time. Dye movement was recorded using time-lapse photography with a digital SLR camera (Nikon D5000 series). 

To determine the experimental concentration profile shown in Fig.~\ref{fig:fig6}, one Hele-Shaw experiment at $\mathrm{Ra} = 120\pm13$ was destructively sampled. After two months of operation, the reservoir fluid was first dyed with rhodamine and then fluorescein, to visualize the downwelling (dark, rhodamine) and upwelling (light, fluorescein) plumes. Once the dynamics of the plumes became apparent, the wet bead packing in the cell was removed in layers, while sampling every 2 cm in depth along the centers of the plumes. The resulting $\sim$1 ml samples were analyzed for their salt concentration using the same protocol as described for the field samples in Appendix~\ref{app:salinity}.

\begin{table*}
    \centering
    \scalebox{0.9}{
    \begin{tabular}{ccccccc}
         grain size & $d_s$ [$\mu$m] & $\kappa$ [m$^2$] & $\varphi$ & $E$ [ms$^{-1}$] & Ra & convect\\
         \hline
         0 – 20 & $1.9\pm 0.2$ & $(1.5 \pm 0.3) \times 10^{-15}$ & 0.37 & $(3.6 \pm 1.0) \times 10^{-7}$ & $0.007\pm 0.002$ & no \\
         70 – 110 & $86\pm 9$ & $(0.8 \pm 0.1) \times 10^{-11}$ & 0.32 & $(6.1 \pm 2.4) \times 10^{-7}$ & $20\pm 9$ & yes \\
         90 – 150 & $123\pm 12$ & $(1.4 \pm 0.1) \times 10^{-11}$ & 0.30 & $(3.6 \pm 1.0) \times 10^{-7}$ & $64\pm 19$ & yes \\
         100 – 200 & $150\pm 15$ & $(1.7 \pm 0.1) \times 10^{-11}$ & 0.28 & $(3.6 \pm 1.0) \times 10^{-7}$ & $74\pm 22$ & yes \\
         150 – 250 & $214\pm 20$ & $(4.7 \pm 0.2) \times 10^{-11}$ & 0.24 & $(3.6 \pm 1.0) \times 10^{-7}$ & $207\pm 59$ & yes \\
         200 – 300 & $264\pm 25$ & $(8.3 \pm 0.3) \times 10^{-11}$ & 0.21 & $(1.0 \pm 0.3) \times 10^{-7}$ & $1298\pm 300$ & yes \\
         \hline
    \end{tabular}
    }
    \caption{Details of experimental tests of the onset of convection, including: grain sizes (from manufacturer) of the glass beads used, along with the Sauter diameter $d_s$, permeability $\kappa$, porosity $\varphi$, evaporation rage $E$ and Rayleigh number Ra.}
    \label{tab:tab3}
\end{table*}

\begin{table*}
 \begin{tabular}{l|c|c|c|c|c|c|c}
  Site  & $d_S$ [$\mu$m] & $\kappa\times 10^{-11}$ [m$^2$] & Ra [Ra$_L$; Ra$_U$] & $\lambda$ [m] & Latitude & Longitude & Year \\
  \hline
  \multicolumn{8}{l}{Death Valley} \\
  \hline
  Badwater Basin (1) 	   & $59\pm15$   & $4.99\pm2.95$  & 28643 [8374; 93961]     & $1.42\pm0.58$ & 36$^\circ$13.651$'$ & -116$^\circ$46.723$'$ & 2016 \\
  Badwater Basin (2)   	   & $66\pm12$   & $6.24\pm2.91$  & 35793 [12736; 103415]   & $1.42\pm0.58$ & 36$^\circ$13.674$'$ & -116$^\circ$46.735$'$ & 2016 \\
  Badwater Basin (3) 	   & $67\pm9$    & $6.52\pm2.56$  & 37444 [15096; 98635]    & $1.27\pm0.55$ & 36$^\circ$13.665$'$ & -116$^\circ$46.820$'$ & 2016 \\ 
  Badwater Basin (4) 	   & $72\pm3$    & $7.39\pm2.29$  & 42419 [20975; 93764]    & $0.58\pm0.32$ & 36$^\circ$13.660$'$ & -116$^\circ$46.903$'$ & 2016 \\ 
  Badwater Basin (5) 	   & $50\pm15$   & $3.10\pm2.35$  & 21064 [5692; 72316]     & $0.55\pm0.28$ & 36$^\circ$13.654$'$ & -116$^\circ$47.036$'$ & 2016 \\
  \hline
  \multicolumn{8}{l}{Owens Lake} \\
  \hline
  T10-3 	           	   & $4.3\pm0.6$ & $0.03\pm0.02$   & 117 [48; 286]          & $1.79\pm0.86$ & 36$^\circ$23.147$'$ & -117$^\circ$56.772$'$ & 2018 \\
  T16 	                   & $6.8\pm1.6$ & $0.07\pm0.04$   & 294 [99; 828]          & $1.19\pm0.51$ & 36$^\circ$23.953$'$ & -117$^\circ$56.454$'$ & 2018 \\
  T2-4 		           	   & $29\pm14$   & $1.23\pm1.12$   & 5457 [918; 21627]      & $1.13\pm0.54$ & 36$^\circ$20.803$'$ & -117$^\circ$58.642$'$ & 2016 \\
  T2-5 (1) 	 	   		   & $24\pm4$    & $0.81\pm0.35$   & 3594 [1443; 8953]      & $1.04\pm0.41$ & 36$^\circ$21.055$'$ & -117$^\circ$58.824$'$ & 2016 \\
  T2-5 (2) 	           	   & $20\pm3$    & $0.55\pm0.22$   & 2436 [1052; 5742]      & $0.94\pm0.50$ & 36$^\circ$20.895$'$ & -117$^\circ$58.740$'$ & 2016 \\ 
  T2-5 (3) 	 	   		   & $18\pm3$    & $0.46\pm0.18$   & 2033 [886; 4762]       & $1.62\pm0.65$ & 36$^\circ$20.877$'$ & -117$^\circ$58.711 & 2018 \\ 

  T25-3 (1) 	           & $11\pm2$    & $0.17\pm0.08$   & 771 [300; 1967]        & $2.25\pm0.89$ & 36$^\circ$27.039$'$ & -117$^\circ$54.510$'$ & 2018 \\
  T25-3 (2) 	           & $24\pm4$    & $0.83\pm0.36$   & 3697 [1484; 9211]      & $1.18\pm0.56$ & 36$^\circ$28.383$'$ & -117$^\circ$54.957$'$ & 2018 \\
  T27-A (1) 	           & $31\pm8$    & $1.35\pm0.81$   & 5994 [1843; 17825]     & $1.70\pm0.65$ & 36$^\circ$29.302$'$ & -117$^\circ$55.953$'$ & 2016 \\
  T27-A (2) 	           & $33\pm4$    & $1.55\pm0.56$   & 6916 [3149; 15614]     & $2.72\pm0.98$ & 36$^\circ$29.061$'$ & -117$^\circ$55.602$'$ & 2016 \\ 
  T27-A (3) 	           & $22\pm3$    & $0.71\pm0.26$   & 3172 [1453; 7126]      & $1.44\pm0.55$ & 36$^\circ$29.112$'$ & -117$^\circ$55.804$'$ & 2018 \\
  T27S $(*)$               & $21\pm6$    & $0.65\pm0.42$   & 2892 [836; 8913]       & $1.51\pm0.64$ & 36$^\circ$28.549$'$ & -117$^\circ$54.994$'$ & 2018 \\
  T29-3 (1) 	           & $138\pm23$  & $27.42\pm12.20$ & 121991 [47476; 310939] & $3.02\pm1.40$ & 36$^\circ$29.955$'$ & -117$^\circ$55.999$'$ & 2016 \\
  T29-3 (2) 	           & $116\pm22$  & $19.49\pm9.29$  & 86704 [32068; 229077]  & $2.80\pm1.34$ & 36$^\circ$29.960$'$ & -117$^\circ$55.962$'$ & 2016 \\

  T32-1-L1 (1) $(*)$       & $36\pm12$   & $1.84\pm1.35$   & 8188 [2018; 27469]     & $1.56\pm0.66$ & 36$^\circ$53.897$'$ & -117$^\circ$57.209$'$ & 2016 \\ 
  T32-1-L1 (2) $(*)$       & $20\pm4$    & $0.56\pm0.26$   & 2492 [946; 6467]       & $2.65\pm0.98$ & 36$^\circ$32.354$'$ & -117$^\circ$57.218$'$ & 2018 \\ 
  T32-1-L1 (3) $(*)$       & $37\pm10$   & $2.01\pm1.18$   & 8923 [2784; 26303]     & $2.43\pm0.92$ & 36$^\circ$32.337$'$ & -117$^\circ$57.204$'$ & 2018 \\ 
  T36-3 (1) 	           & $46\pm4$    & $3.04\pm1.01$   & 13538 [6483; 29282]    & $1.17\pm0.91$ & 36$^\circ$29.953$'$ & -117$^\circ$58.505$'$ & 2016 \\
  T36-3 (2) 	           & $35\pm8$    & $1.82\pm0.99$   & 8086 [2689; 22918]     & $2.27\pm1.03$ & 36$^\circ$30.050$'$ & -117$^\circ$58.518$'$ & 2016 \\ 
  T36-3 (3) 	           & $75\pm25$   & $8.08\pm5.83$   & 35924 [9027; 119336]   & $1.43\pm0.62$ & 36$^\circ$29.724$'$ & -117$^\circ$57.916$'$ & 2016 \\ 
  T8-W 	 	           	   & $8.7\pm2.6$ & $0.11\pm0.08$   & 485 [135; 1527]        & $0.87\pm0.41$ & 36$^\circ$22.522$'$ & -117$^\circ$57.256$'$ & 2018 \\
  \hline
  \multicolumn{8}{l}{Sua Pan} \\
  \hline
  B7 		   & $9.6\pm0.9$  & $0.13\pm0.05$   & 341 [117; 1683] & $0.92\pm0.48$ & -20$^\circ$35.046$'$ & 25$^\circ$54.654$'$ & 2012 \\ 
  L5		   & $11.6\pm1.1$ & $0.20\pm0.08$   & 497 [170; 2456] & $0.59\pm0.22$ & -20$^\circ$33.996$'$ & 26$^\circ$0.420$'$ & 2012 \\
  D10 		   & $10.6\pm1.4$ & $0.16\pm0.07$   & 415 [131; 2198] & $0.41\pm0.16$ & -20$^\circ$36.678$'$ & 25$^\circ$55.794$'$ & 2012 \\
  J11 		   & $10.6\pm1.4$ & $0.16\pm0.07$   & 415 [131; 2198] & $0.69\pm0.27$ & -20$^\circ$37.242$'$ & 25$^\circ$59.250$'$ & 2012 \\
  B3 		   & $10.6\pm1.4$ & $0.16\pm0.07$   & 415 [131; 2198] & $0.58\pm0.22$ & -20$^\circ$32.880$'$ & 25$^\circ$54.672$'$ & 2012 \\
  D5 		   & $10.6\pm1.4$ & $0.16\pm0.07$   & 415 [131; 2198] & $0.95\pm0.33$ & -20$^\circ$33.972$'$ & 25$^\circ$55.818$'$ & 2012 \\
  I4 		   & $10.6\pm1.4$ & $0.16\pm0.07$   & 415 [131; 2198] & $0.53\pm0.20$ & -20$^\circ$33.447$'$ & 25$^\circ$58.699$'$ & 2012 \\ 
  \hline
 \end{tabular}
  \caption{\label{tab:tab4}Sauter diameter $d_S$, permeability $\kappa$, Rayleigh number Ra, lower Ra limit Ra$_L$, upper Ra limit $Ra_U$ and pattern wavelength $\lambda$ measured or calculated for each of the field sites at Badwater Basin, Owens Lake and Sua Pan. For each site, longitude, latitude and year of the field campaign are also given. Samples stemming from a trench are indicated as $(*)$.}
\end{table*}

\clearpage

%

\end{document}